\documentclass[aip,
 amsmath,amssymb,
 reprint,%
]{revtex4-1}

\usepackage{graphicx}%
\usepackage{dcolumn}
\usepackage{amsmath}
\usepackage{color}
\usepackage{bm}
\usepackage[utf8]{inputenc}
\usepackage[T1]{fontenc}
\usepackage{mathptmx}
\usepackage{etoolbox}

\makeatletter
\def\@email#1#2{%
 \endgroup
 \patchcmd{\titleblock@produce}
  {\frontmatter@RRAPformat}
  {\frontmatter@RRAPformat{\produce@RRAP{*#1\href{mailto:#2}{#2}}}\frontmatter@RRAPformat}
  {}{}
}%
\makeatother

\begin{document}

\title[{\it Perspective on $\mu$SR under hydrostatic pressure}]{Perspective on muon-spin rotation/relaxation under hydrostatic pressure}
%
\author{Rustem Khasanov}
 \email{rustem.khasanov@psi.ch}
 \affiliation{Laboratory for Muon Spin Spectroscopy, Paul Scherrer Institut, CH-5232 Villigen PSI, Switzerland}

\begin{abstract}
Pressure, together with temperature, electric, and magnetic fields, alters the system and allows for the investigation of the fundamental properties of matter.
Under applied pressure, the interatomic distances shrink, which modifies the interactions between atoms and may lead to the appearance of new (sometimes exotic) physical properties such as: pressure-induced phase transitions; quantum critical points; new structural, magnetic and/or superconducting states; changes of the temperature evolution and symmetry of the order parameters; and so on.

Muon-spin rotation/relaxation ($\mu$SR) has proven to be a powerful technique in elucidating the magnetic and superconducting responses of various materials under extreme conditions. At present,  $\mu$SR experiments may be performed in high magnetic field up to $\sim 9$~T, temperatures down to $\simeq 10-15$~mK and hydrostatic pressure up to $\sim 2.8$~GPa.
In the following Perspective paper, the requirements for $\mu$SR experiments under pressure, the existing high-pressure muon facility at the Paul Scherrer Institute (Switzerland), and selected experimental results obtained by $\mu$SR under pressure are discussed.

\end{abstract}


\maketitle

\section{Introduction}

Muon-spin rotation/relaxation ($\mu$SR) experiments provide information on the atomic level about the chemical and physical properties of matter. Compared to other microscopic techniques, $\mu$SR experiments are relatively easy and rather straightforward to perform. Indeed, in $\mu$SR studies:
\begin{itemize}
  \item a relatively complicated sample environment can be used as illustrated by the large number of measurements performed  down to 10~mK in temperature, up to 9~T in magnetic field, up to 2.8~GPa under hydrostatic pressure {\it etc.};
  \item collection of a single measurement point typically requires less than an hour. In other words, in a reasonable amount of time it is possible to obtain the $\mu$SR response of the sample and follow the evolution as a function of temperature, magnetic field, pressure {\it etc.};
  \item the sample may be in various aggregate states as a gas, liquid, or solid;
  \item the sample does not need to contain any specific nuclei as is required in NMR and NQR experiments.
\end{itemize}

High pressure research has become one of the most significant areas in $\mu$SR studies as it allows for the fine tuning of different phases of matter, as well as to follow their strong interaction/interplay.
The importance of $\mu$SR studies under pressure led to the development of dedicated muon instrumentation.\cite{Butz_HI_1986, Kratzer_HI_1994, Andreica_thesis_2001, Watanabe_PhysB_2009, Khasanov_HPR_2016, Shermadini_HPR_2017, Naumov_PRA_2011, Khasanov_HPR_2022}
In this Perspective paper we concentrate on the particular application of the muon-spin rotation/relaxation technique for conducting experiments under hydrostatic pressure conditions. Particular attention is paid to the description of the existing high-pressure muon facility at the Paul Scherrer Institute (PSI), Switzerland.

The paper is organized as follows: section \ref{Sec:Basic-Principles} comprises a short description of the $\mu$SR method. Particular attention is paid to the combination of hydrostatic pressure and $\mu$SR techniques (Sec.~\ref{Sec:Muon-Stopping_Sites}). The muon implantation depth and requirements for the $\mu$SR pressure cell construction are discussed in Sec.~\ref{Sec:Implantation_Depth}.  The recent developments in $\mu$SR pressure instrumentation are presented in Sec.~\ref{Sec:Instrumentation}. These include the description of the $\mu$E1 decay beamline (with particular attention dedicated to the formation of the spin-rotated muon beam, Sec.~\ref{Sec:Decay_beam-lime}); the upgraded version of the General Purpose Decay (GPD) muon spectrometer (Sec.~\ref{Sec:GPD_spectrometer}); the construction of the three-wall piston-cylinder pressure cell (Sec.~\ref{Sec:Three-wall_cell});  and the optical setup for the in-situ pressure determination (Sec.~\ref{Sec:Optical_Setup}). A few scientific examples demonstrating the versatility and ability of $\mu$SR under pressure to study magnetism, superconductivity, and the rich interplay between these two phenomena are discussed in Sec.~\ref{Sec:Examples}.

\begin{figure*}[htb]
\centering
\includegraphics[width=0.8\linewidth, angle=0]{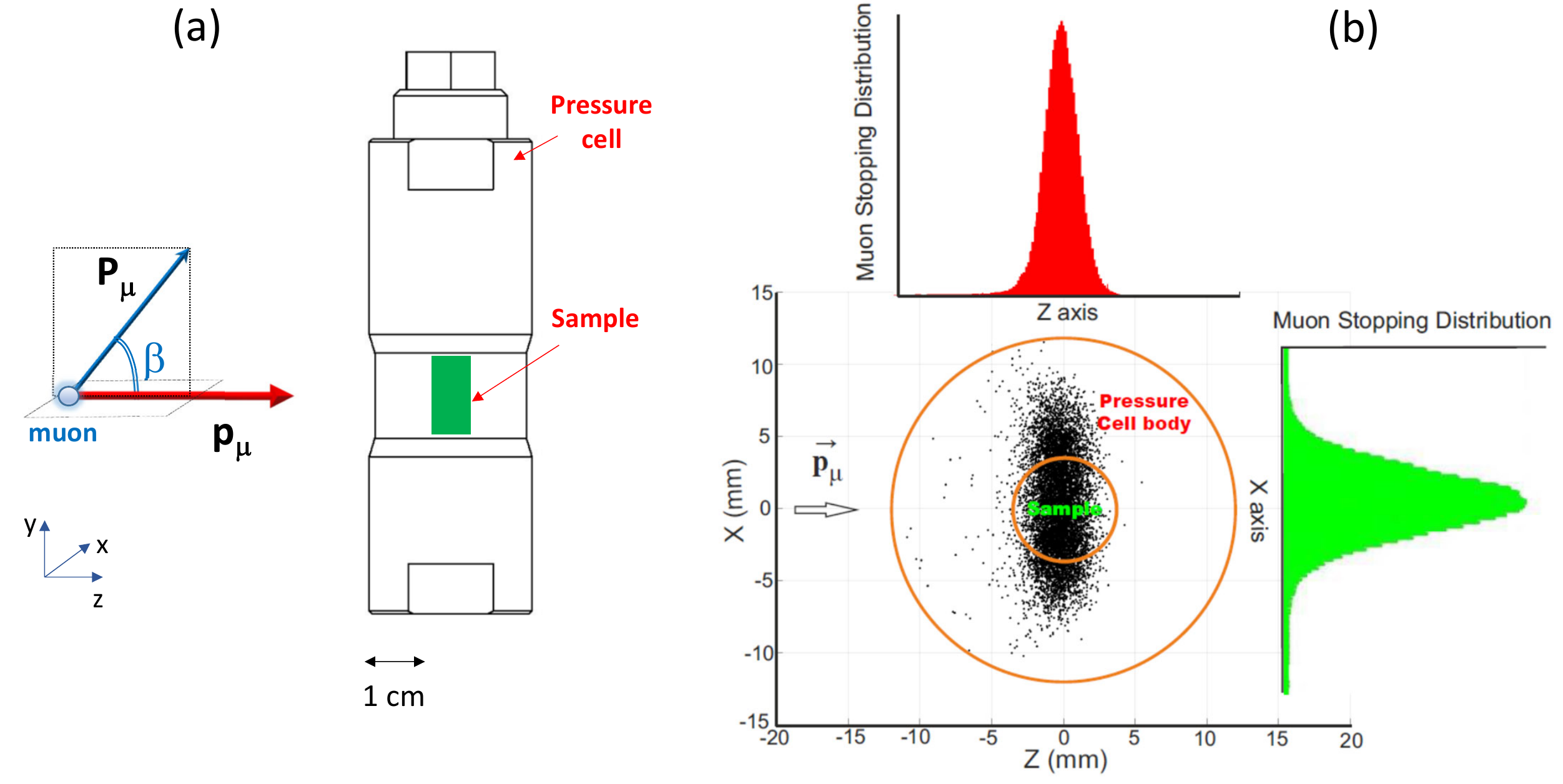}
\caption{(a) Schematic view of a cylindrical pressure cell (black contour) with the sample inside (green rectangle). The muons are implanted along the $z-$axis. The spin of the incoming muons may be rotated by an angle $\beta$ from the direction of the muon momentum ${\bf p}_\mu$. (b) The cross sectional view ($x-z$) of the pressure
cell. The colored areas represent the muon stopping distributions in parallel (red) and perpendicular (green) directions to the muon beam. The
energy of the implanted muons was set to 44 MeV. The simulations were performed using the TRIM.SP package, Ref.~\onlinecite{TRIM-SP}. The figure is adapted after Ref.~\onlinecite{Shermadini_HPR_2017}. }
 \label{Fig:muon-stopping}
\end{figure*}

\section{Basic principles of $\mu$SR experiments under pressure}\label{Sec:Basic-Principles}

This Section discusses the basic principles of $\mu$SR experiments under pressure. The Section starts from the calculated muon stopping distribution inside a typical $\mu$SR pressure cell. The scattering of muons at the pressure cell walls sets limitations for the energy of the implanted muons as well as for the construction of the $\mu$SR pressure cells.

\subsection{The muon stopping inside the pressure cell}\label{Sec:Muon-Stopping_Sites}

A schematic view of the pressure $\mu$SR experimental setup is presented in Figure~\ref{Fig:muon-stopping}. The muons are implanted along the $z-$axis. The initial spin of the incoming muons ({\it i.e.} the initial muon-spin polarization ${\bf P}_\mu$) may stay nearly parallel (or antiparallel) to the muon momentum ${\bf p}_\mu$, or rotated by an angle $\beta$
(see also Sec.~\ref{Sec:Decay-muon_Production} for different types of spin-rotated muons accessible at the decay muon beamlines).

$\mu$SR experiments may be conducted in different configurations, which are distinguished by an angle between the applied magnetic field and the initial muon-spin polarization of the incoming muon beam, as well as without any applied field.
A detailed description of the zero-, transverse- and longitudinal-field $\mu$SR experiments may be found in the following: textbooks, in Refs.~\onlinecite{Schenck_book_1985, Smilga_book_1994, Schenk_book_1995, Karlsson_book_1995, Lee_book_1999, Yaouanc_book_2011, Blundell_book_2022}; and review articles, in Refs.~\onlinecite{Blundell_ContPhys_1999, Bakule_ContPhys_2004, Amato_RMP_1977, Dalmas_JPCM_1997, Sonier_RMP_2000, Khasanov_HPR_2016, Khasanov_SST_2015, Hillier_NaturePrimer_2022}.

Upon entering the pressure cell, muons are scattered at the cell walls, and when reaching the sample, the muon beam diverges in both parallel and perpendicular directions with respect to the muon beam. The cross sectional view of muon spread inside the pressure cell walls and the sample is presented in Figure~\ref{Fig:muon-stopping}~(b). The simulations were performed using the TRIM.SP package for a pressure cell made of the MP35N alloy (the outer and the sample diameters 24 and 6~mm, respectively) and for a muon beam width of 4~mm.  The energy of the implanted muons was set to 44~MeV.

\subsection{Muon implantation energy and pressure cell construction}\label{Sec:Implantation_Depth}

The results presented in Fig.~\ref{Fig:muon-stopping} set limitations to the energy of the implanted muons as well as to possible constructions of the $\mu$SR pressure cell.
Indeed, by approaching the sample, the muon beam diverges quite substantially. The FWHM of the muon stopping distribution reaches $\simeq 4$ and 8~mm in the directions parallel and perpendicular to the beam, respectively.
Consequently, in order to stop a substantial amount of muons inside the sample ({\it i.e.} not in the pressure cell walls), the sample volume must be relatively high (of the order of few hundred cubic millimeters). At the same time, the pressure cell needs to be compact enough to fit inside the cryogenic environment and within the muon detector block. By following the 'maximum pressure' {\it vs.} 'sample volume' diagram presented by Klotz in Ref.~\onlinecite{Klotz_book_2013}, it becomes obvious that only the piston-cylinder pressure cell construction satisfies such criteria.

\begin{figure}[htb]
\centering
\includegraphics[width=1.0\linewidth, angle=0]{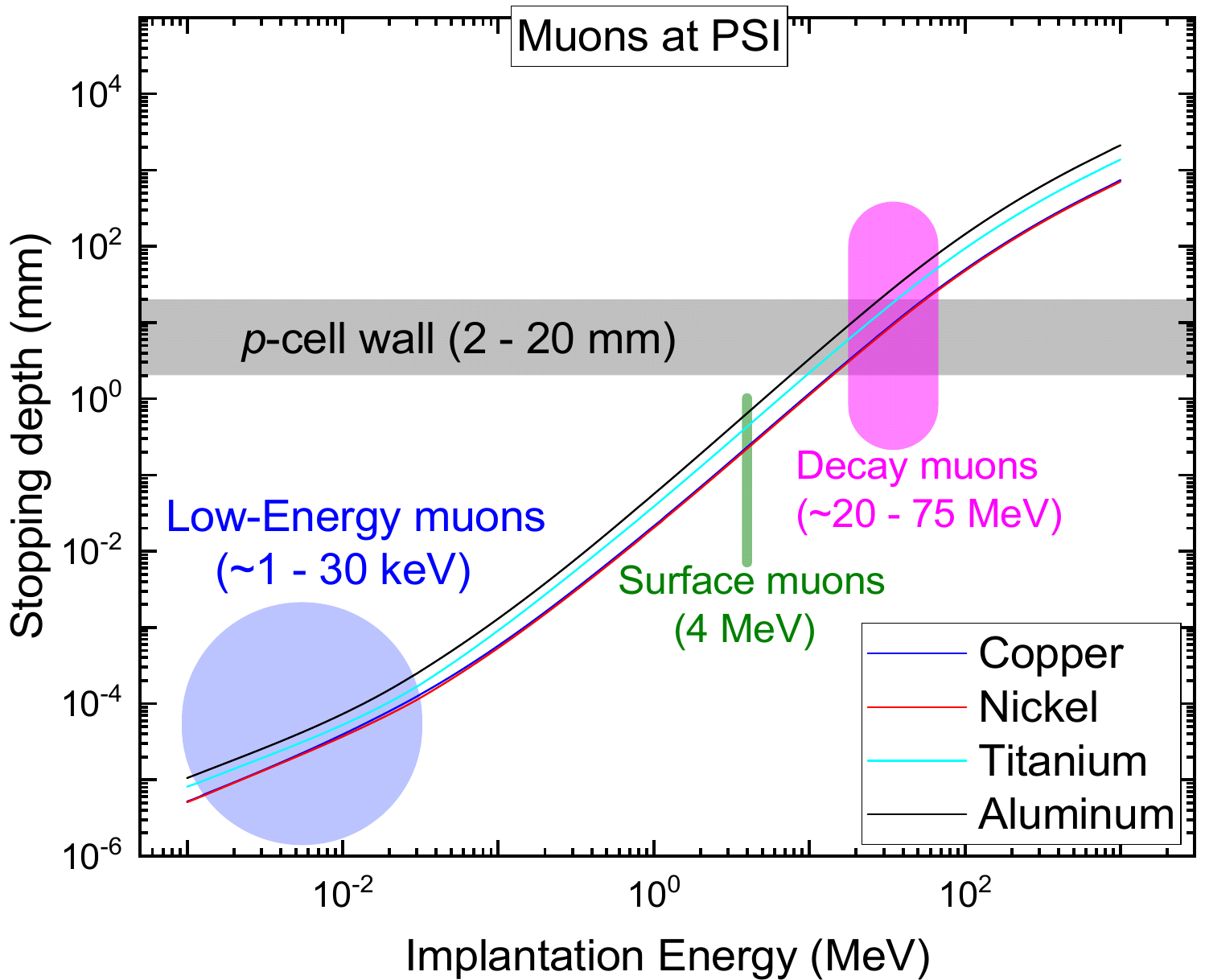}
\caption{Dependence of the mean stopping depth on the implantation energy for muons produced at various beam facilities at the Paul Scherrer institute. The calculations were performed by using TRIM.SP package, Ref.~\onlinecite{TRIM-SP}. }
 \label{Fig:Implantation_Energy}
\end{figure}

As for the energy of the implanted muons.  Figure~\ref{Fig:Implantation_Energy} presents the dependence of the mean stopping depth as a function of the implantation energy for muons produced at various beamline facilities at the Paul Scherrer Institute (PSI). The calculations were performed for four different elemental metals, namely copper, nickel, titanium and aluminum. These metals are either used as materials for the cells (as, {\it e.g.}, aluminum), or enter various hard alloys as the main component (see, {\it e.g.}, Refs.~\onlinecite{Klotz_book_2013, Khasanov_HPR_2016, Khasanov_HPR_2022, Shermadini_HPR_2017, Walker_RSI_99, Uwatoko_JPCM_2002} and Table~\ref{Table:Alloys} in Sec.~\ref{Sec:construction-materials}). The horizontal grey stripe corresponds to the pressure wall thickness ranging from 2 to 20~mm. Obviously, only the so called 'decay-muons' with an implantation energy ranging between $\simeq 20$ and 75~MeV are suitable for $\mu$SR studies under pressure.

\section{Instrumentation}\label{Sec:Instrumentation}

This Section provides an overview of the instrumentation used to perform $\mu$SR experiments under hydrostatic pressure at  the Paul Scherrer Institute, Switzerland. It includes the description of the $\mu$E1 decay beamline (with particular attention paid to the formation of the spin-rotated muon beam), the upgraded version of the General Purpose Decay (GPD) muon spectrometer, the construction of the three-wall piston-cylinder pressure cell, and the optical setup for the in-situ pressure determination.
The description of the old version of the GPD spectrometer, the double-wall piston cylinder pressure cells, and the method of pressure determination by means of AC-susceptibility may be found in Refs.~\onlinecite{Khasanov_HPR_2016, Shermadini_HPR_2017, Naumov_PRA_2011}.

\subsection{The $\mu$E1 decay muon beamline}\label{Sec:Decay_beam-lime}

\subsubsection{The decay muon production}\label{Sec:Decay-muon_Production}

In large scale facilities such as PSI,  the muon beam is produced by highly energetic protons. Protons are collided into a target made of carbon, thus creating a cascade of pions which further decay into muons. The two step process of producing the positively charged muon beam can be described by:
\begin{equation}
{\rm p}+{\rm p}\rightarrow {\rm p}+{\rm n}+\pi^+ \nonumber
 \label{eq:protop-pion}
\end{equation}
for the pion and
\begin{equation}
\pi^+\rightarrow \mu^+ + \gamma_{\mu}\nonumber
 \label{eq:pion-muon}
\end{equation}
for the muon production, respectively. Here ${\rm p}$, $\pi$, $\mu$, ${\rm n}$ and $\gamma_\mu$ denote the proton, pion, muon, neutron, and muon neutrino, respectively. The spin of the pion is zero, however the muon neutrino has definite helicity. This requires that, within the reference frame of the pion, the spin of the positively charged muon is aligned {\it opposite} the muon momentum.\cite{Schenck_book_1985, Smilga_book_1994, Schenk_book_1995, Karlsson_book_1995, Lee_book_1999, Yaouanc_book_2011, Blundell_book_2022}
Large-scale facilities can provide two types of muon beams, namely 'surface' and 'decay' beams. A description of surface muon beamlines can be found in textbooks Refs.~\onlinecite{Schenck_book_1985, Smilga_book_1994, Schenk_book_1995, Karlsson_book_1995, Lee_book_1999, Yaouanc_book_2011, Blundell_book_2022}, as well as in review articlesRefs.~\onlinecite{Blundell_ContPhys_1999, Bakule_ContPhys_2004, Amato_RMP_1977, Dalmas_JPCM_1997, Sonier_RMP_2000}.

\begin{figure}[htb]
\centering
\includegraphics[width=1.0\linewidth, angle=0]{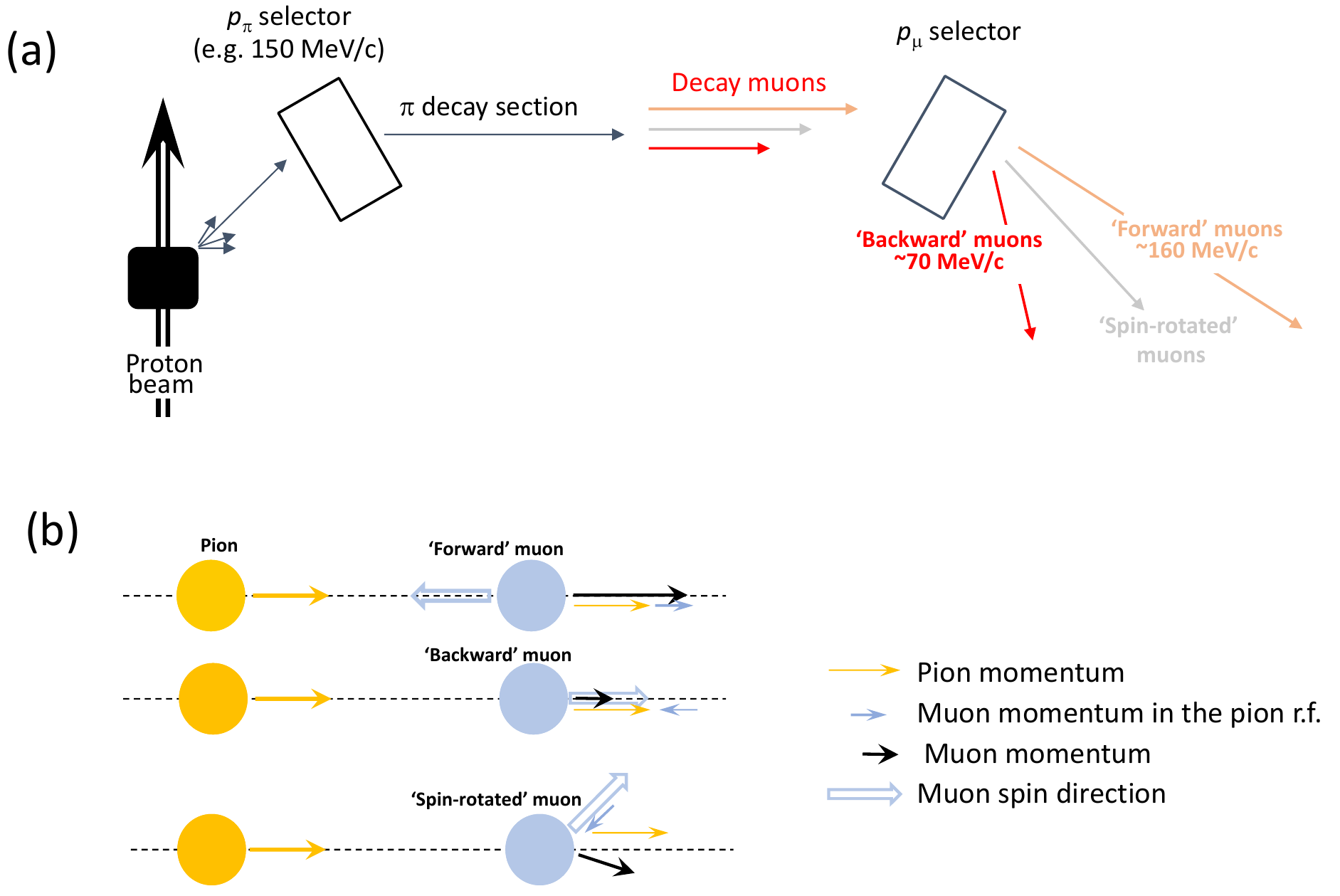}
\caption{(a) A schematic representation of the decay muon beamline. The high-energy pions are ejected from the target and transported down to a pion momentum selection element ('$p_\pi$ selector'). The pion decay occurs in flight inside the pion decay channel ('$\pi$ decay section').  The muon momentum selection part ['$p_{\mu}$ selector'] allows to choose muons with the desired energy. (b) Three different types of decay muons which might be collected at the end of the pion decay section. The 'Forward' muons have the highest possible momentum and the muon-spin aligned opposite to the muon momentum. The 'Backward' muons have the smallest momentum and the spin aligned along the muon momentum. The 'Spin-rotated' muons are emitted at a certain angle from the beamline axis. The spin direction of such muons may, in principle, vary between 0 and 180 degrees. }
 \label{Fig:Decay_beam-line}
\end{figure}

Figure~\ref{Fig:Decay_beam-line}~(a) explains the working principles of a decay beamline. The high-energy pions are ejected from the target and transported down to a pion momentum selection element ('$p_\pi$ selector'). The pion decay occurs in-flight inside the pion decay channel ('$\pi$ decay section'). It is important that the length of the pion's trajectory within the decay channel should be comparable with the pion decay length. The muons, which are captured at the end of the pion decay section, might be further transported to the spectrometer. The muon momentum selection  ['$p_{\mu}$ selector'] allows to choose muons with the desired energy.

It is important to emphasize here that a decay beamline allows one to obtain muon beams with different initial muon-spin polarizations. This is different from a surface beamline, where the muon-spin and the muon momentum are aligned opposing each other (here we do not consider the presence of a Wien-filter-based spin-rotation device).
As is explained in Fig.~\ref{Fig:Decay_beam-line}~(b), the particular muon-spin polarization at the decay beamline  depends on the muon emission direction within the reference frame of the pion.
In practice, two extreme conditions are commonly used:
\begin{itemize}
  \item [1.]  The muon is emitted in the direction of the pion momentum, {\it i.e.} in the 'forward' direction [the top diagram in Fig.~\ref{Fig:Decay_beam-line}~(b)]. The resulting muon momentum (black arrow) consists of the muon momentum in the reference frame of the pion (blue thin arrow) and the momentum of the pion (yellow arrow). The final muon momentum of the 'forward' muon is, therefore, greater than the one of the pion and the muon-spin points in the opposite direction of its propagation.
  \item [2.] The muon is emitted in the direction opposite to the pion momentum, {\it i.e.} in the 'backward' direction [the middle diagram in Fig.~\ref{Fig:Decay_beam-line}~(b)]. The final muon momentum is smaller than that of the pion. The muon spin is aligned along the muon momentum.
  \end{itemize}
The intermediate case, {\it i.e.} muons with the muon-spin rotated at a certain angle from the main beam axis [the bottom diagram at Fig.~\ref{Fig:Decay_beam-line}~(b)] was not in use until recently. This is an important configuration, however, since it allows $\mu$SR experiments to be performed in the so-called spin-rotated mode. In such a case:
\begin{itemize}
  \item [3.]  The final momentum of the 'spin-rotated' muon becomes a vector sum of the pion and the muon momentum in the pion's reference frame [the lowest diagram in Fig.~\ref{Fig:Decay_beam-line}~(b)]. The spin of the muon is aligned opposite to the muon momentum in the reference frame of the pion and it stays, therefore, turned (rotated) at a certain angle from the beam trajectory. The absolute value of the muon momentum may stay between values allowed for the 'forward' and 'backward' muon, respectively. 
  \end{itemize}

It is important to emphasize, that within the 'spin-rotated' configuration the muon is emitted at an angle to the beam direction [see the black arrow at the lowest diagram of Fig.~\ref{Fig:Decay_beam-line}~(b)], so one needs certain modification of the beam transport in order to bring such muons back to the main trajectory. The first successful attempt of using the spin-rotated muons at a decay beamline was made at the M9B beamline in TRIUMF, Canada.\cite{Goko_private-comm}

\subsubsection{$\mu$E1 beamline with asymmetrically driven quadrupole magnet} \label{Sec:mue1}

\begin{figure}[htb]
\centering
\includegraphics[width=1.0\linewidth, angle=0]{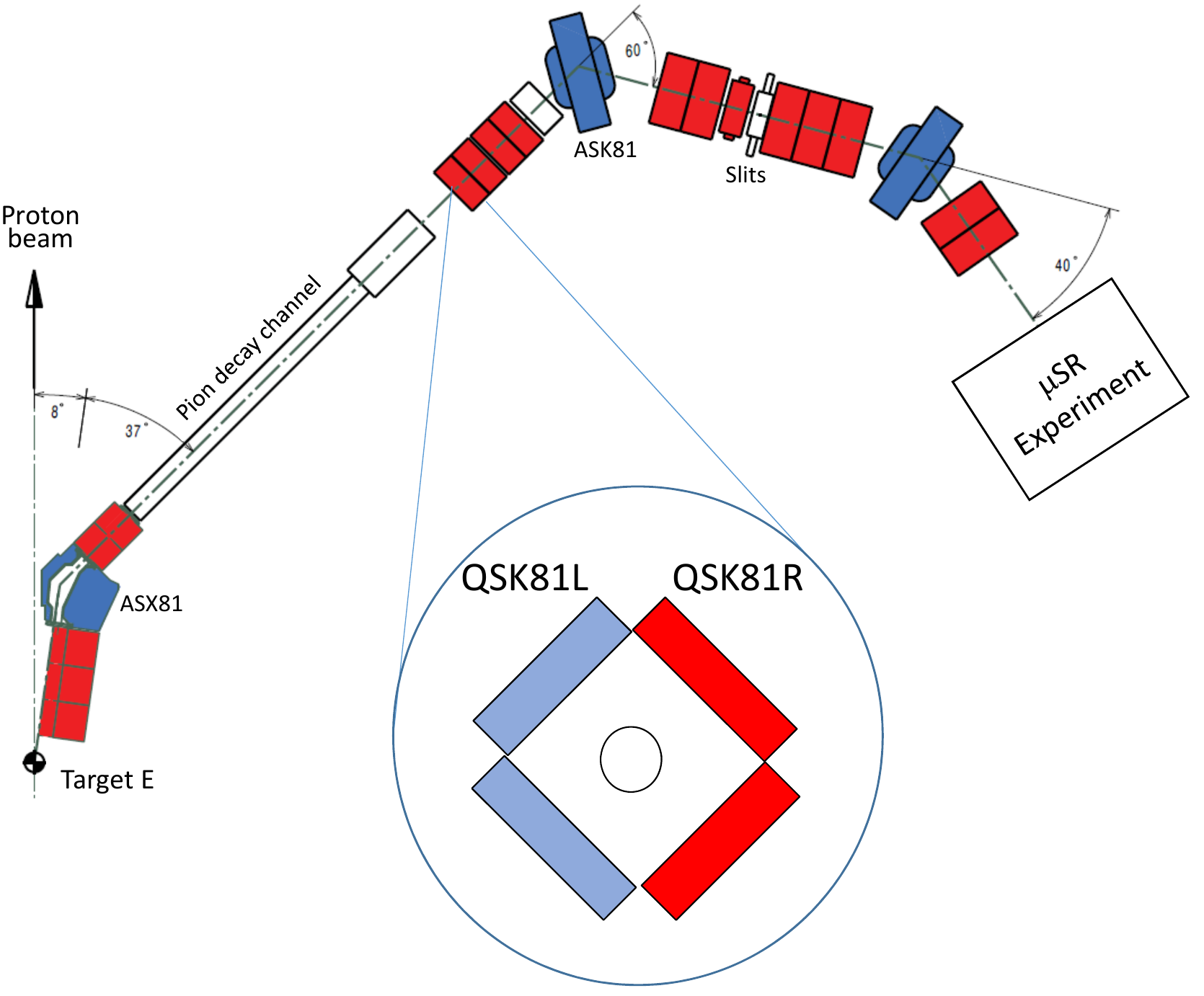}
\caption{A schematic view of the $\mu$E1 decay muon beamline at the Paul Scherrer Institute,  Switzerland. The 'beam optics' consist of magnets used to transport the muon beam to the measuring station. The magnets are either quadrupoles (red rectangles) or bending magnets (blue elements). Slits are used to limit the lateral beam extension and reduce the beam intensity. The extension inside the circle represents the asymmetrically driven QSK81 quadrupole magnet. QSK81L and QSK81R stand for the 'left' and 'right' coils which are run by separate power supplies.  }
 \label{Fig:mue1}
\end{figure}

The $\mu$E1 decay muon beamline installed at PSI was previously discussed in Ref.~\onlinecite{Khasanov_HPR_2016}. The schematic view of the beamline is presented in Fig.~\ref{Fig:mue1}. The 'beam optics' consist of various magnets used to transport the muon beam to the experimental measuring station. All magnets are tunable, thus allowing for precise control of both the beam momentum and the beam spread. The magnets are either quadrupoles (red rectangles) or bending magnets (blue elements).   The quadrupoles, which stay in a sequence of 2, 3 or 4 elements, are used to focus the muon beam, while the bending magnets act as momentum selection elements. Slits  are used to limit the lateral beam extension and reduce the beam intensity.

The transport of the pion/muon beam occurs in the following way. The high-energy pions leave the 'Target E' at high energies. They are collected over a certain solid angle by quadrupole magnets. The bending magnet ASX81 acts as a pion momentum selector. The pion decay section consists of an 8~m long superconducting solenoid with a field of $\simeq 5$~T. Pions decay before reaching the end of the solenoid, so the end of the solenoid section becomes a diffuse source of the muons. The selection of the muon momentum is performed at the ASK81 bending magnet. The rest of the beam elements are used to transport muons to the experimental station.

All quadrupole magnets except QSK81 are run 'symmetrically', {\it i.e.} currents flowing through all four magnet coils stay equal. This is the standard operational mode of a quadrupole magnet.\cite{Quadrupole_Wiki}  In order to collect the 'spin-rotated' muons [see Sec.~\ref{Sec:Decay-muon_Production} and Fig.~\ref{Fig:Decay_beam-line}~(b)] connections to QSK81, {\it i.e.} the first quadrupole magnet at the end of the pion decay channel, were modified. The 'left' (QSK81L) and the 'right' (QSK81R) pair of coils were run from separate power supplies (see the schematic representation of the magnet in Fig.~\ref{Fig:mue1}). In such a configuration, the QSK81 magnet may operate in both 'symmetric' and 'asymmetric' modes.

\subsubsection{The muon-spin rotation at $\mu$E1 beamline }

The $\mu$E1 beamline, in its original design presented in Ref.~\onlinecite{Khasanov_HPR_2016}, is capable to transport the 'forward' and the 'backward' muons [the top and the intermediate panels in Fig.~\ref{Fig:Decay_beam-line}~(b)]. In such configurations, the muon-spin is aligned either antiparallel or parallel to the muon momentum and it remains, therefore, 'non-rotated'.
%
%
In order to collect the 'spin-rotated' muons, the first beam focusing element at the end of the pion decay section must be able to compensate for deviation of the muon momentum from the beamline axis [see the lowest panel at Fig.~\ref{Fig:Decay_beam-line}~(b)]. It requires the QSK81 magnet to have both quadruplolar (focusing) and dipolar (bending) components. This becomes possible by running the quadrupolar QSK81 magnet in the aforementioned  'asymmetric' mode (see Sec.~\ref{Sec:mue1} and Fig.~\ref{Fig:mue1}).

\begin{figure}[htb]
\centering
\includegraphics[width=1.0\linewidth, angle=0]{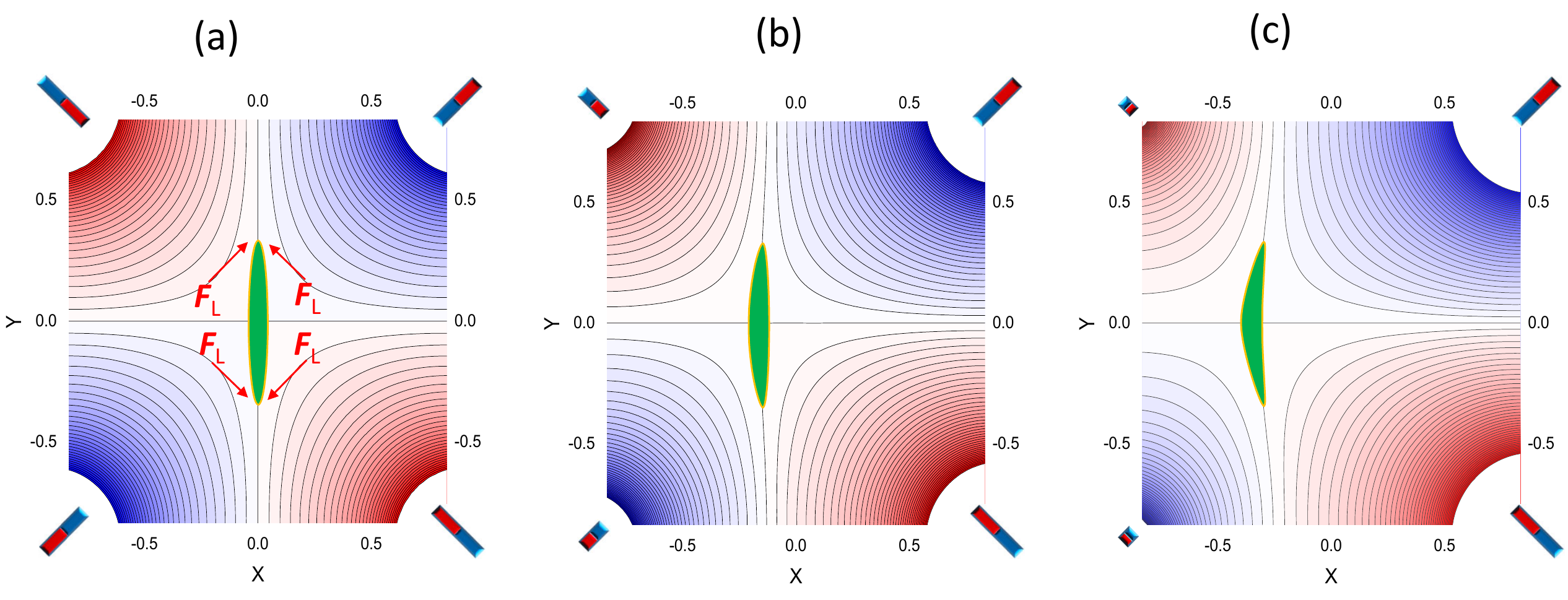}
\caption{(a) The equipotential magnetic field lines of a symmetrically driven quadrupole magnet. Arrows indicate directions of the Lorenz force. The green oval at the center represents the beam spot. The quadrupole focuses the beam horizontally and de-focuses it in the vertical direction. (b) and (c) The equipotential magnetic field lines of a quadrupole magnet in the asymmetric mode. The left pair of magnets is weaker compared to the right one. The appearance of the dipolar component, {\it i.e.} the shift of the beam spot from the central position in horizontal direction, is clearly visible.  }
 \label{Fig:quadrupole-simulations}
\end{figure}

The simulations of an asymmetrically driven quadrupole magnet are presented in Fig.~\ref{Fig:quadrupole-simulations}. Here, the poles of the quadrupole are modeled by four permanent magnets placed at corners. The solid lines represent the equipotential field surfaces. As expected, with all magnets being equal ({\it i.e.} in the symmetric case), the quadrupole focuses the beam of charged particles in one direction (horizontally) and de-focuses it in another direction (vertically), see Fig.~\ref{Fig:quadrupole-simulations}~(a). By decreasing the magnetic field strengths of one pair of magnets [left pair in our case, Figs.~\ref{Fig:quadrupole-simulations}~(b) and (c)], the beam spot slightly distorts and shifts from the central position. Both the dipolar and quadrupolar components are clearly visible. One may also note the presence of the higher order components (sextuple, octuple {\it etc.}) which lead to small additional distortions of the beam-spot. These components, however, have a secondary importance and they are not considered here.

\begin{figure}[htb]
\centering
\includegraphics[width=1.0\linewidth, angle=0]{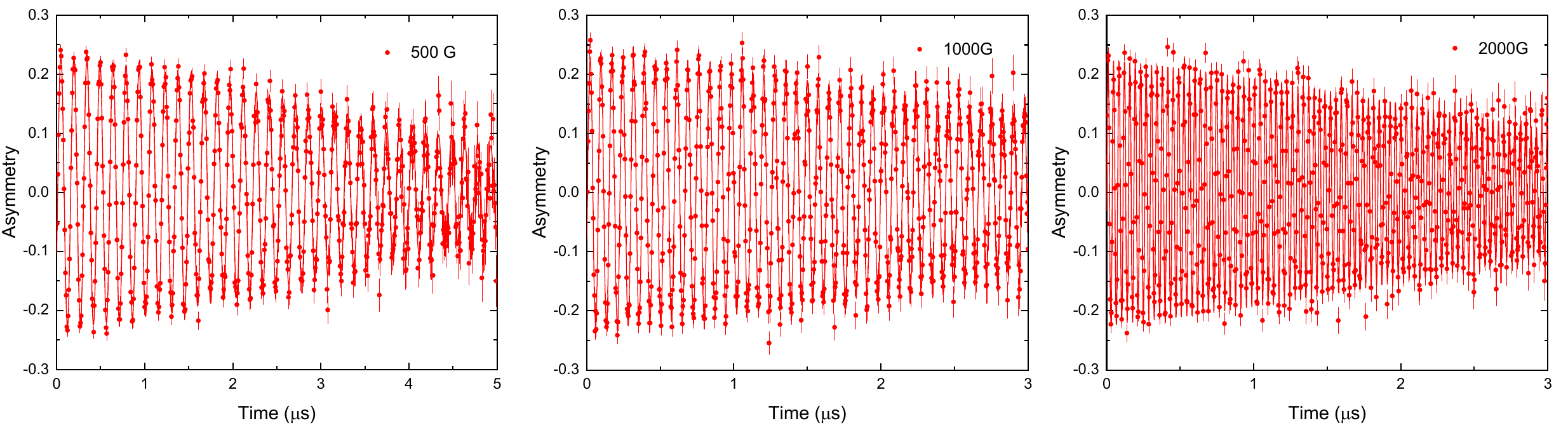}
\caption{(a) The muon time spectra collected using the spin-rotated muon beam at the $\mu$E1 beamline. The external magnetic field $\mu_0H=0.05$~T is applied parallel to the muon momentum [along the $z-$axis in Fig.~\ref{Fig:muon-stopping}~(a)]. The muon momentum is kept at 100~meV/c. (b) and (c) The same as in panel (a) but for $\mu_0H=0.1$~T and 0.2~T, respectively. }
 \label{Fig:TF_time-spectra}
\end{figure}

The resulting transversal-field $\mu$SR time spectra, accumulated after adjustment of the $\mu$E1 beamline for the 'spin-rotated' muon transport,  are presented in Fig.~\ref{Fig:TF_time-spectra}. In these experiments the magnetic field 0.05~T [panel (a)], 0.1~T [panel (b)], and 0.2~T [panel (c)] was applied along the muon momentum [along the $z-$axis in Fig.~\ref{Fig:muon-stopping}~(a)]. The muon-spin rotation angle was estimated to be $\beta\simeq 60^{\rm o}$. Note that this angle is comparable with the best muon-spin rotation device installed at the $\pi$M3 surface muon beamline at PSI.\cite{Amato_RMP_1977}

At the end of this Section, the following three important points need to be mentioned:
\begin{itemize}
  \item Decay beamlines are capable to collect and transport the 'spin-rotated' muons to the measuring station. The amount of spin-rotated muons transported to the sample at the $\mu$E1 beamline at PSI is high enough to perform $\mu$SR experiments in the time-differential mode ($\sim 20\cdot10^3$ muons per second per 1~mA proton current down to the muon momentum $p_{\mu}\simeq 90$~MeV/c).
  \item The major element allowing to capture the spin-rotated muons is the asymmetrically driven quadrupole, where both the quadrupolar (focusing) and the dipolar (bending) components are present.
  \item The rotation of the muon-spin occurs without the use of the so-called Wien-filter setup as is performed {\it e.g.}, at the 'surface' and low-energy muon beamlines.\cite{Amato_RSI_2017, Salman_PhysProcedia_2012}
\end{itemize}.

\subsection{The General Purpose Decay (GPD) $\mu$SR spectrometer}\label{Sec:GPD_spectrometer}

The GPD (General Purpose Decay) $\mu$SR spectrometer is permanently installed at the end of the $\mu$E1 beamline (see Sec.~\ref{Sec:mue1} and Fig.~\ref{Fig:mue1}). The previous version of the GPD muon instrument was described in Ref.~\onlinecite{Khasanov_HPR_2016}. The detector was built using photomultiplier tubes and plastic light guides.

\begin{figure*}[htb]
\includegraphics[width=1.0\linewidth]{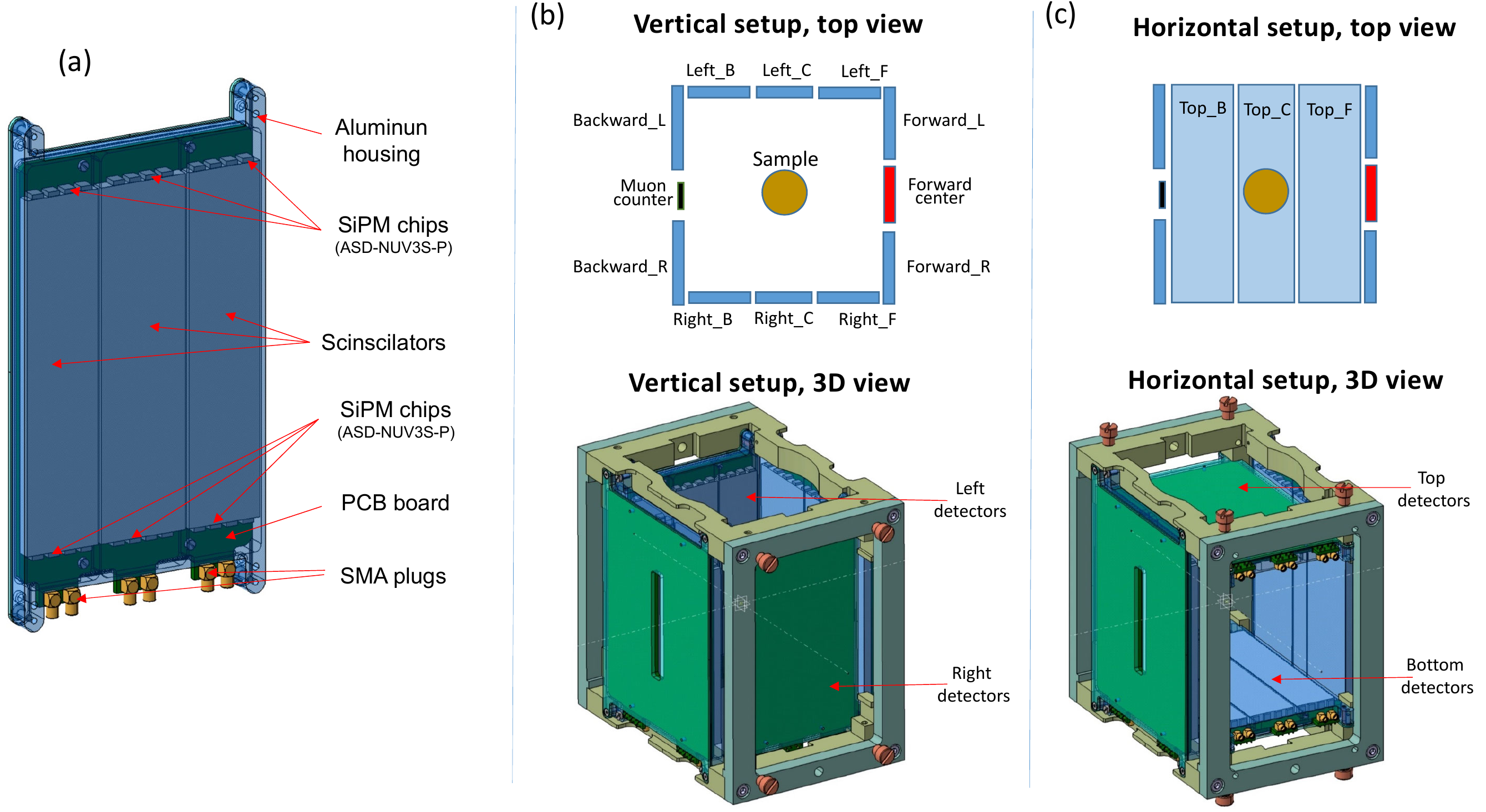}
\caption{(a) Construction of a SiPM-based detector unit. The full block consists of three detectors. For each individual detector, the light from the scintillator is read out on two opposite sites by arrays of 4 SiPM's (ASD-NUV3S-P) mounted on a printed electronics board and glued directly onto the scintillator material. (b) The 'vertical setup' of the GPD detector head. (b) The 'horizontal setup' of the GPD detector head. The change between the 'vertical' and the 'horizontal' setups is done by remounting the 'Left' and 'Right' detector units onto the 'Top' and the 'Bottom' positions, respectively}
 \label{Fig:GPD_detectors}
\end{figure*}

The new GPD detectors have been constructed using the silicon photomultipliers (SiPM's). The SiPM based $\mu$SR detectors deliver performances similar
to that of photomultiplier tubes, Refs~\onlinecite{Stoykov_PhysicaB_2009, Sedlak_PhysicaB_2009, Sedlak_PhysicaB_2009_2, Stoykov_PhysicsProcedia_2012, LEM_SiPM, Amato_RSI_2017}, but  they are also compact and insensitive to magnetic fields. This allows the SiPM chips to be mounted directly on the scintillator portion of the detector [see, {\it e.g.}, Ref.~\onlinecite{Amato_RSI_2017} and Fig.~\ref{Fig:GPD_detectors}~(a)]. The SiPM-based $\mu$SR detectors were first developed to equip the avoided level-crossing instrument (ALC).\cite{Stoykov_PhysicaB_2009, Sedlak_PhysicaB_2009} Later on, similarly designed detectors were used at the high magnetic field instrument HAL-9500,\cite{Sedlak_PhysicaB_2009_2, Stoykov_PhysicsProcedia_2012} the low-energy muon instrument (LEM),\cite{LEM_SiPM} and  the General Purpose Surface (GPS) spectrometer,\cite{Amato_RSI_2017} all of which are found at PSI.

Prior to the upgrade of the GPD instrument, the design and efficiency of a new detector system were simulated using the $musrSim$ and $musrSimAna$ software packages.\cite{Sedlak_PhysProc_2012} These software packages were sucessfully used to simulate several $\mu$SR instruments.\cite{Amato_RSI_2017, Sedlak_PhysicaB_2009_2, Sedlak_IEEE_2010} The $musrSim$ program, based on the Geant4,\cite{Agostinelli_PhysResA_2003, Allison_IEEE_2006} and Root packages,\cite{Brun_PhyResA_1997}  calculates the detector response to the muons and their decay products. The output of $musrSim$ was further analysed with the general $\mu$SR analysis tool, $musrSimAna$. The subsequent use of these packages allows to implement the full logic of a real $\mu$SR experiment as the coincidences and anti-coincidences between different detectors, as well as to obtain the time-independent background of the detector histograms. The reliability of the simulations was further tested by comparing the measured and simulated parameter values.

Figure~\ref{Fig:GPD_detectors}~(a) demonstrates the construction of one SiPM-based GPD detector unit. The single block, placed inside the light-tightened Aluminum housing, consists of three detectors. For each individual detector, the light from the scintillator is read out on two opposite sites by arrays of several SiPM's (ASD-NUV3S-P from AdvanSiD, Ref.~\onlinecite{Advansid}) mounted on a printed electronics board and glued directly onto the scintillator material. The SiPMs are powered by a PSI-designed multi-channel power supply (PSI HVR800). The analog signals from the SiPM is amplified by broadband amplifiers.\cite{Stoykov_PhysicsProcedia_2012, Cattaneo_IEE_2014} The output of the amplifiers is processed by constant fraction discriminators (PSI CFD-950, Ref.~\onlinecite{Prokscha_PhysicaB_2009}) and the signals are finally sent to a time-to-digital converter (CAEN, TDC V1190B). The event time for the incoming positron(muon) is calculated from the average time between the events recorded on both opposite edges.

Four detector units are screwed together by forming a 'detector head'. Figures~\ref{Fig:GPD_detectors}~(b) and (c) show two possible layouts of the GPD detector head, which differ in terms of cryostat access direction and detector logic.  The 'vertical setup' is used for cryostats loaded from the top [Fig.~\ref{Fig:GPD_detectors}~(b)], while the 'horizontal setup' allows access from the side [Fig.~\ref{Fig:GPD_detectors}~(c)]. The change between the setups is made by remounting the 'Left' and 'Right' detector units onto the 'Top' and the 'Bottom' positions, respectively [see the lower panels at Figs.~\ref{Fig:GPD_detectors}~(b) and (c)].

As for the detector logic.  In both, the 'vertical' and 'horizontal', configurations the 'Backward' and the 'Forward' outputs are formed similarly. The positrons accumulated at the 'Backward\_L' and 'Backward\_R' detectors are summed together by forming a single 'Backward' output: ${\rm Backward}={\rm Backward\_L}+{\rm Backward\_R}$.  The 'Forward' output is formed by either coupling all three detectors (${\rm Forward}={\rm Forward\_L}+{\rm Forward \  center}+{\rm Forward\_R}$), or by coupling only two of them (${\rm Forward}={\rm Forward\_L}+{\rm Forward\_R}$) and leaving the 'Forward center' as a so-called veto counter. Note that the veto mode is generally used to reject the muons missing the sample (see, {\it e.g.}, Ref.~\onlinecite{Amato_RSI_2017} for explanation of the veto principle).

As for the 'Left'/'Top' and 'Right'/'Bottom' set of detectors.  Within the 'vertical setup,' the coupling occurs for the opposite sections of the 'Left' and 'Right' detector units. The corresponding output channels are: ${\rm Backward\_LR}={\rm Left\_B}+{\rm Right\_B}$, ${\rm Center\_LR}={\rm Left\_C}+{\rm Right\_C}$ and ${\rm Forward\_LR}={\rm Left\_F}+{\rm Right\_F}$, respectively.
Note that, due to the highly asymmetric emission of decay positrons, the 'Backward\_LR' and 'Forward\_LR' channels account for nearly 60-65\% of the initial muon-spin polarization and, because of this, they may be involved in the data analysis process.
Within the 'Horizontal setup,' the individual detectors in the 'Top' and 'Bottom' sections are coupled together by forming large top (${\rm Top}={\rm Top\_B}+{\rm Top\_C}+{\rm Top\_F}$) and bottom (${\rm Bottom}={\rm Bottom\_B}+{\rm Bottom\_C}+{\rm Bottom\_F}$) positron counters, respectively.

\begin{figure}[htb]
\centering
\includegraphics[width=1.0\linewidth, angle=0]{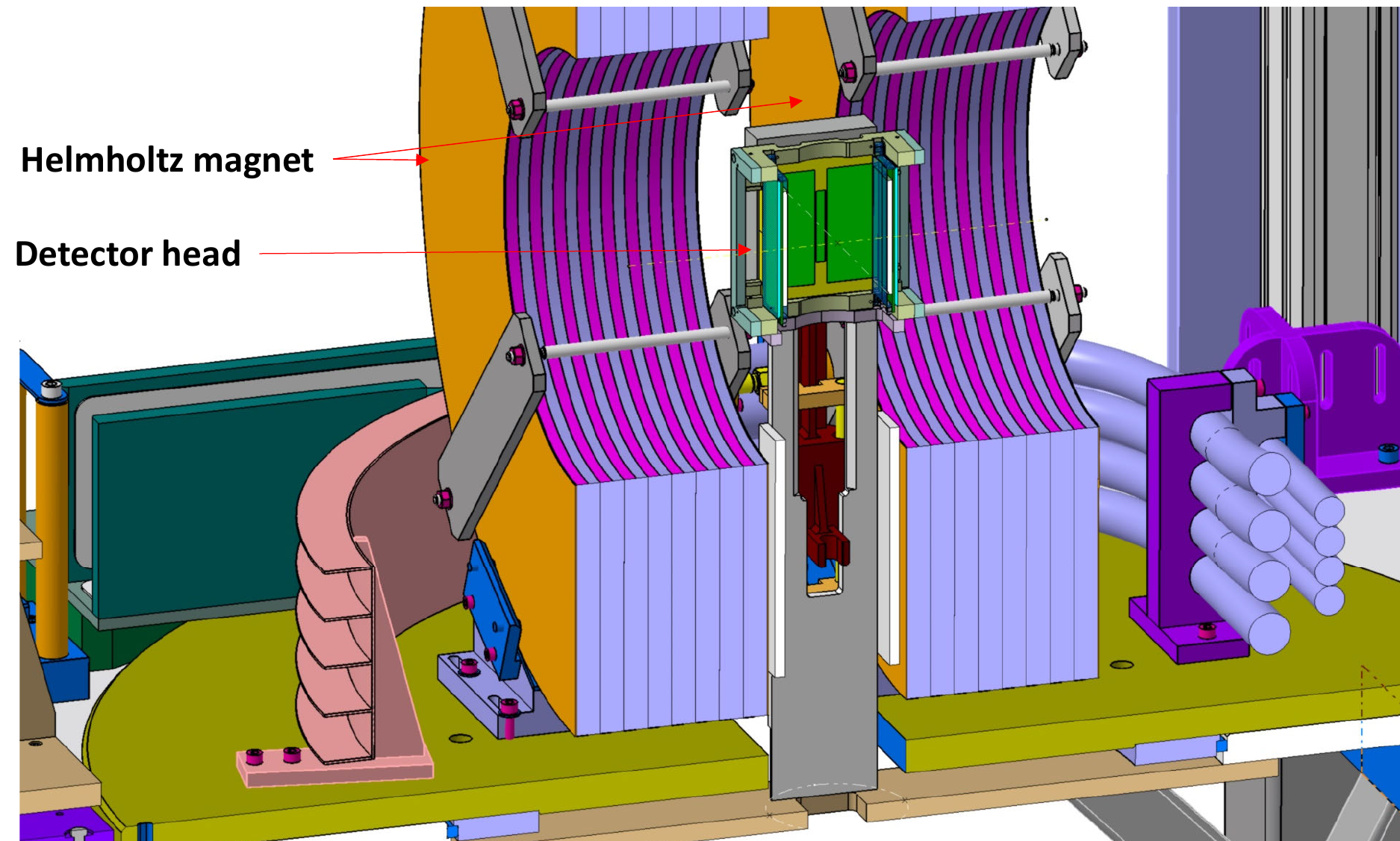}
\caption{The SiPM based detector head mounted on the GPD instrument platform. }
 \label{Fig:GPD}
\end{figure}

The dimensions of the detector head (with all the detector units screwed together) is approximately  12x14x15~cm$^3$.  Figure \ref{Fig:GPD} demonstrates the mounting of the detector head inside the Helmholz magnet, which is fixed to the GPD platform (see also Ref.~\onlinecite{Khasanov_HPR_2016} describing the construction of the old version of the GPD instrument).

\subsection{$\mu$SR pressure cells}

In this Section the construction materials, design of the three-walled $\mu$SR pressure cell, and the optical setup for determining the pressure inside the piston-cylinder cells are discussed. The single and double-wall $\mu$SR pressure cell construction, as well as a way of pressure determination by means of AC-susceptibility were presented in Refs.~\onlinecite{Shermadini_HPR_2017, Khasanov_HPR_2016, Grinenko_SRO_NatCom_2021, Andreica_thesis_2001}.

\subsubsection{The construction materials}\label{Sec:construction-materials}

Upon entering the cell muons are scattered at the pressure cell walls such that the beam has diverged in both transversal and longitudinal directions by the time it reaches the sample. The simulations presented in Fig.~\ref{Fig:muon-stopping}~(b) imply that a substantial amount of muons stop in the pressure cell walls. This requires that the construction materials, which are used to build the cells, must have a low or, at best, magnetic-field- and temperature-independent $\mu$SR response. They should also be nonmagnetic, otherwise the induced magnetism in the cell walls may affect the sample response.

Materials used to build piston-cylinder pressure cells should be divided into materials suitable for the pressure cell body and for the compressive pistons. Obviously, the body of the cell should be able to withstand both compressive and tensile deformations, while pistons should be able to resist very high compressions.

Table~\ref{Table:Alloys} summarizes the list of nonmagnetic alloys, which may be used to build the bodies of $\mu$SR cells. For such materials the most relevant parameters are the yield strength (the maximum strength up to which material responds elastically, $\sigma_{\rm Y}$), the Young modulus (the slope of the linear part of the stress-strain curve for a material under tension/compression, $E$), and ductility (a measure of a material's ability to undergo significant deformation before rupture or breaking, $\Delta l/l$). The rest of the mechanical parameters, as well the mechanical and thermal treatment procedures, can be found at the corresponding producer's web sites and some dedicated papers and books focused on the pressure technique (see, {\it e.g.}, Ref.~\onlinecite{Klotz_book_2013, Walker_RSI_99, Uwatoko_JPCM_2002}).

\begin{table}[htb]
\caption{\label{Table:Alloys} The yield strengths ($\sigma_{\rm Y}$), Young modulus ($E$) and ductility ($\Delta l/l$) for some nonmagnetic alloys, which could be used to build $\mu$SR pressure cell bodies. The values of $\sigma_{\rm Y}$, $E$, and $\Delta l/l$ are given for $T\simeq 300$~K. The meaning of the alloys: Al-7075 -- aluminum alloy; TAV6 -- titanium alloy containing 90 wt.\% Ti, 4 wt.\% Al, and 6 wt.\% V; CuBe -- beryllium copper (Berylco-25); NiCrAl -- nickel based alloy containing 57 wt.\% Ni, 40 wt.\% Cr, and 3 wt.\% Al; MP35N -- Ni/Co based alloy containing 35 wt.\% Ni, 35 wt.\% Co, 20 wt.\% Cr, and 10 wt.\% Mo.  All materials are aged to the optimum conditions allowing to get maximum possible value of the Yield strengths.\cite{Klotz_book_2013, Walker_RSI_99, Uwatoko_JPCM_2002}}

\begin{center}
\begin{tabular}{lcccccccc}\\
 \hline
 \hline
&Al-7075&TAV6&CuBe&NiCrAl&MP35N\\
 \hline
$\sigma_{\rm Y}$~(GPa) &0.5& 1.05 & 1.3 &2.06  & 2.15  \\
$E$~(GPa) &71.7&97 & 131& 190& 215 \\
$\Delta l/l$~(\%)&7&12&5&4-5&9 \\
 \hline \hline \\
\end{tabular}
   \end{center}
\end{table}

The pistons should withstand high compressive strength, which may allow the use of sintered compounds. The list of materials suitable to for producing nonmagnetic pistons is summarised in Table~\ref{Table:Sinterred_Materials}.

\begin{table}[htb]
\caption{\label{Table:Sinterred_Materials} The compressive strengths ($\sigma_{\rm Y}$) and the Young modulus ($E$) for selected sintered compounds. The values of $\sigma_{\rm Y}$ and  $E$ are given for $T\simeq 300$~K.\cite{Klotz_book_2013, Walker_RSI_99, Uwatoko_JPCM_2002} }
\begin{center}
\begin{tabular}{lcccccccc}\\
 \hline
 \hline
&ZrO$_2$-Y$_2$O$_3$&Al$_2$O$_3$-ZrO$_2$&Si$_3$N$_4$&SiC&WC\\
 \hline
$\sigma_{\rm Y}$~(GPa) &2.2&4.7&5.1&8.3&5.0-11.0  \\
$E$~(GPa)              &210&357&241&918& 600-670 \\
 \hline \hline \\
\end{tabular}
   \end{center}
\end{table}

The $\mu$SR pressure cells used at the Paul Scherrer Institute are made from CuBe, NiCrAl and MP35N nonmagnetic alloys. The pistons are made of Yttrium stabilized zirconia oxide (ZrO$_2$-Y$_2$O$_3$),  silicon nitride (Si$_3$N$_4$), or nickel bound tungsten carbide (WC). The $\mu$SR response of some of the aforementioned alloys and ceramic materials can be found in Refs.~\onlinecite{Khasanov_HPR_2016, Shermadini_HPR_2017, Shermadini_PhD-Thesis_2014, Andreica_thesis_2001}

\subsubsection{Three-wall pressure cell}\label{Sec:Three-wall_cell}

\begin{figure}[htb]
\centering
\includegraphics[width=1.0\linewidth, angle=0]{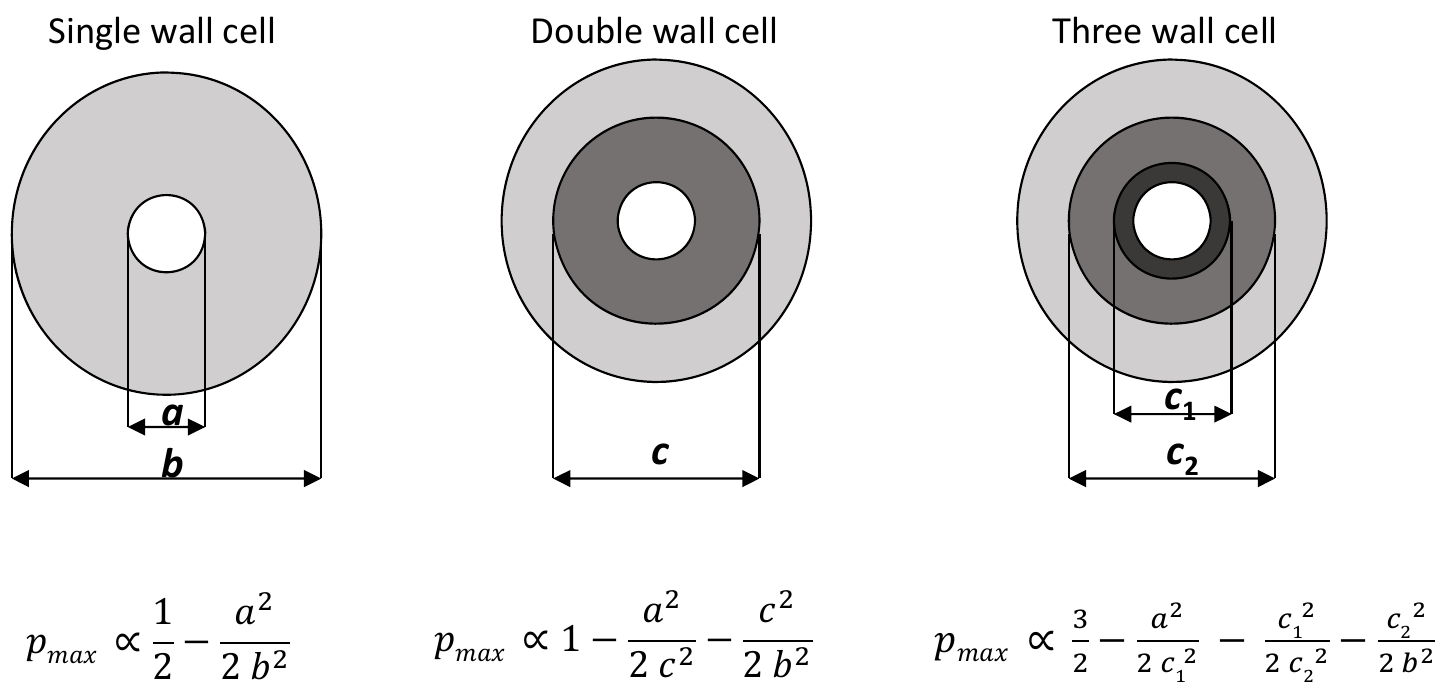}
\caption{The radial cross-section of the single-wall, the double-wall, and the three-wall compound cylinders.
$a$ and $b$ are the inner and the outer diameters of the cylinder assembly. $c$, $c_1$, and $c_2$ are the diameters of the inner cylinders. Equations represent maximum pressures for each particular cylinder geometry.}
 \label{Fig:P-cells}
\end{figure}

One innovative method to increase the maximum achievable pressure in piston-cylinder cells is the use of the so-called compound cylinder design. In this case, the cells are made of two (or more) monobloc cylinders which are shrink-fitted into each other.\cite{Eremets_book_1996, Klotz_book_2013, Uwatoko_JPCM_2002, Walker_RSI_99, Khasanov_HPR_2016, Shermadini_HPR_2017, Khasanov_HPR_2022}
The radial cross section of the single-wall ($s$), the double-wall ($d$), and the three-wall ($t$) cylinders are presented in Fig.~\ref{Fig:P-cells}. Here $a$ and $b$ are the inner and the outer diameters of the cylinder assembly, while $c$, $c_1$, and $c_2$ are the diameters of the inner cylinders. For each particular geometry, the maximum pressure may be determined from the Lam\'{e} equations, see for example Refs.~\onlinecite{Eremets_book_1996, Klotz_book_2013, Shermadini_PhD-Thesis_2014}. The maximum pressure values depend on the mechanical properties of the material (mostly on the yield strengths $\sigma_{\rm Y}$, see Table~\ref{Table:Alloys}), the interface pressure between the cylinders, and the ratios of the cylinder diameters. Considering the optimum interface pressures, and assuming that all cylinders are made of the same materials, the Lam\'{e} equations result in the maximum pressures as presented in the lower part of Fig.~\ref{Fig:P-cells}.

For the compound cylinder with $a=6$~mm and $b=24$~mm, which are typical dimensions for the pressure cells used in $\mu$SR experiments,\cite{Khasanov_HPR_2016, Shermadini_HPR_2017} one gets:\cite{Khasanov_HPR_2022}
\begin{equation}
p_{\rm max}^s \; \div p_{\rm max}^d \; \div p_{\rm max}^t \simeq 1\; \div 1.6\; \div 1.92.
\label{eq:max-pressure_ratios}
\end{equation}
This suggests that the three-wall assembly allows to almost double the maximum pressure compared to the single-wall design and reaches $\simeq20$\% higher values compared to the double-wall geometry.

\begin{figure}[htb]
\centering
\includegraphics[width=1.0\linewidth, angle=0]{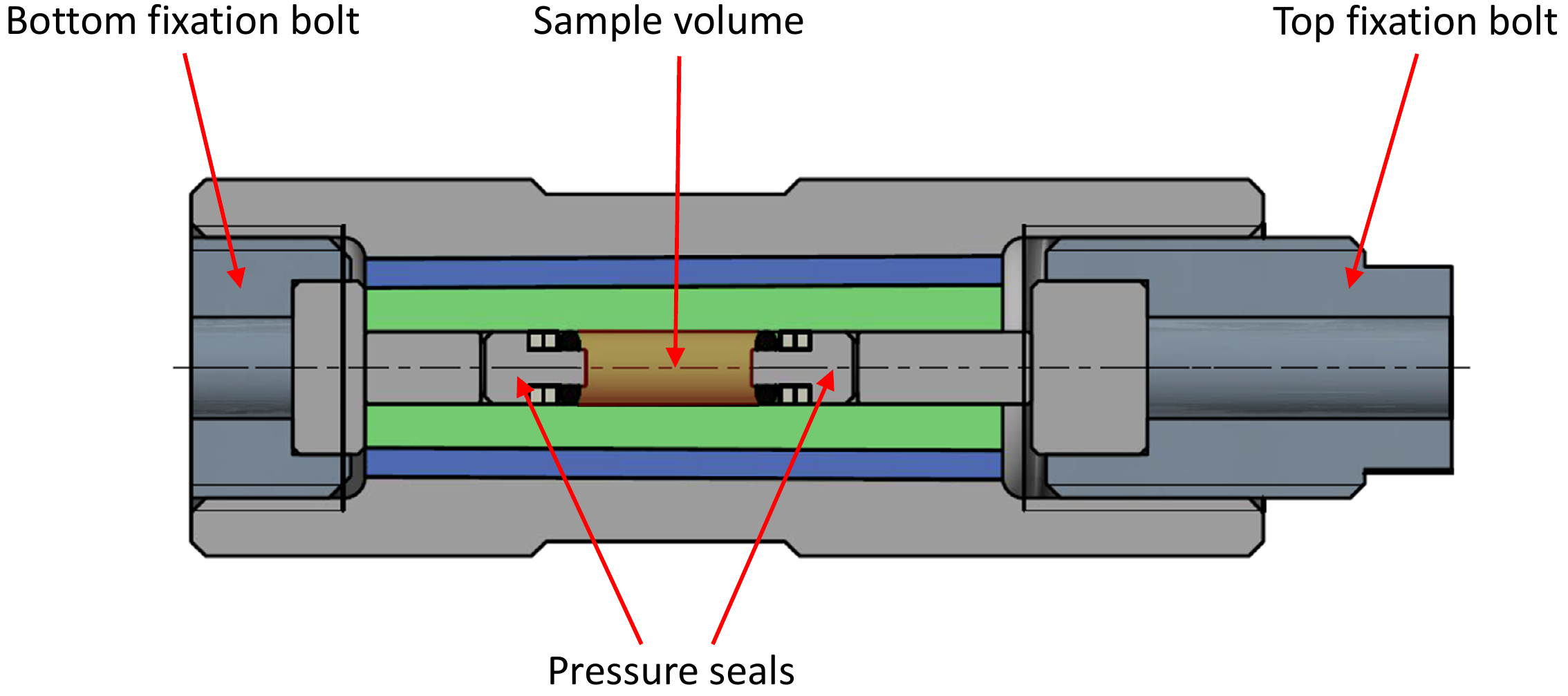}
\caption{The cross-sectional of the three-wall pressure cell. After Ref.~\onlinecite{Khasanov_HPR_2022}. }
 \label{Fig:three-wall-cell}
\end{figure}

The first attempt to built the three-wall pressure cell for $\mu$SR experiments is reported in Ref.~\onlinecite{Khasanov_HPR_2022}. The mechanical drawing of the assembled three-wall pressure cell is shown in Fig.~\ref{Fig:three-wall-cell}. The outer cylinder of the cell body is made from MP35N, while the middle and the inner cylinders were made from NiCrAl nonmagnetic alloys. The mechanical design and performance of the pressure cell were evaluated and optimised using finite-element analysis. The outcome of experimental testing closely matches the modeled results. The three-wall $\mu$SR cell has achieved pressures up to $\simeq3.3$~GPa at ambient temperature, corresponding to $\simeq3.0$~GPa at low temperatures, without irreversible damage.

\subsubsection{Double volume pressure cell for optical pressure determination}\label{Sec:Optical_Setup}

Measurement of the absolute value of the pressure inside the piston-cylinder clamp cells used in $\mu$SR experiments is a complicated task. Until present, the pressure inside the $\mu$SR cell was determined via AC-susceptibility measurements by monitoring the pressure-dependent shift of the superconducting transition temperature ($T_{\rm c}$) of a small piece of elemental metal (In, Sn, or Pb).\cite{Andreica_thesis_2001, Khasanov_HPR_2016, Shermadini_HPR_2017, Khasanov_HPR_2022} This method has some disadvantages since: (i) the piece of the superconducting metal (the pressure indicator) stays in close vicinity to the sample and may affect the sample response, (ii) the pressure indicator occupies part of the sample volume and (iii) the absolute value of pressure is measured at a single temperature, which corresponds to $T_{\rm c}$ of the pressure indicator (below $\sim7$~K).

\begin{figure}[htb]
\centering
\includegraphics[width=1.0\linewidth, angle=0]{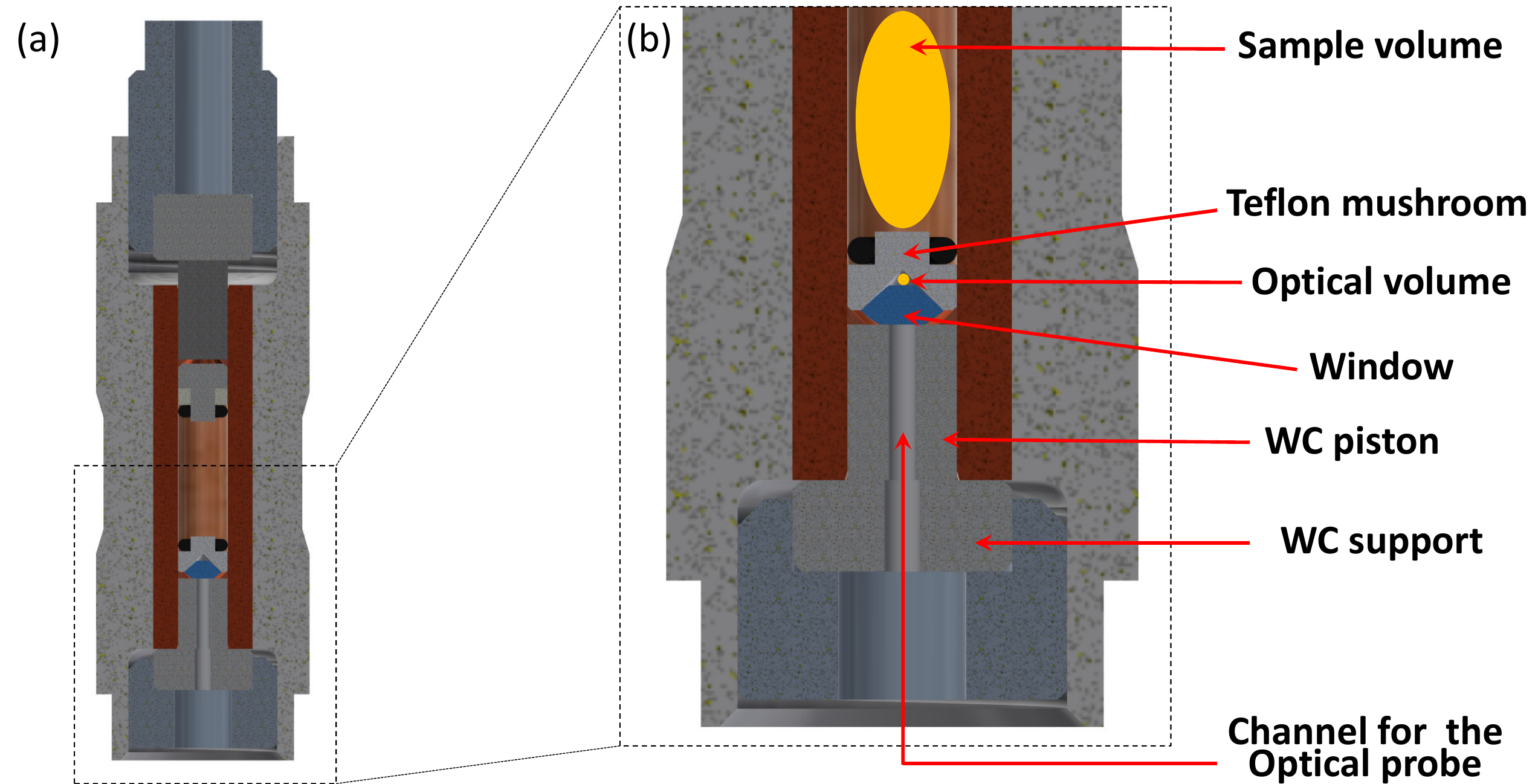}
\caption{(a) The cross-sectional view of the double-volume piston-cylinder pressure cell. The pressure cell body (with the inner and the outer diameters $\varnothing6$ and $\varnothing24$~mm, respectively) and the top part of the pressure seal [the mushroom, the tungsten carbide (WC) piston, the WC support, and the fixation bolt] are the same as described in Refs.~\onlinecite{Khasanov_HPR_2016, Shermadini_HPR_2017, Khasanov_HPR_2022}. (b) The expanded part of the cell with the optical setup. The main elements are: the teflon mushroom with the conical entrance, the optical window prepared from the commercial gem crystal with the `brilliant-cut' shape, and the WC piston with the entrance for optical fibers. The small and large ovals represent the volume available for the sample and the optical probe, respectively. After Ref.~\onlinecite{Naumov_PRA_2011}. }
 \label{Fig:double-volume_pressure-cell}
\end{figure}

In order to monitor the pressure inside the cell without reducing the volume available for the sample, as well as to prevent an influence of the pressure indicator material on the sample space, the concept of a double-volume pressure cell where the 'optical' and the 'sample' volumes are physically separated from each other was developed and successfully tested in Ref.~\onlinecite{Naumov_PRA_2011}. As indicated in Fig.~\ref{Fig:double-volume_pressure-cell}, the construction of the piston-cylinder cell was modified by introducing a tungsten carbide piston with an entrance for optical fibers, a teflon seal with a conical cavity for an optical pressure indicator, and an optical window made of commercially available cubic-Zirconia single crystals. The technique developed in Ref.~\onlinecite{Naumov_PRA_2011} allows the determination of the pressure inside the cell with an accuracy better than $0.01$~GPa and might be potentially used to monitor the pressure inside the cell within the full temperature range allowed at the GPD spectrometer (from $\simeq 0.25$~K up to room temperature).

\section{Scientific Examples}\label{Sec:Examples}

Muon spin rotation/relaxation can be used to study magnetic materials due to its high sensitivity to small fields and capability to probe both the static and dynamic local field distributions. 

Zero-field $\mu$SR is used to investigate the microscopic magnetic properties of solids.\cite{Schenck_book_1985, Smilga_book_1994, Schenk_book_1995, Karlsson_book_1995, Lee_book_1999, Yaouanc_book_2011, Blundell_book_2022, Blundell_ContPhys_1999, Bakule_ContPhys_2004, Amato_RMP_1977, Dalmas_JPCM_1997, Sonier_RMP_2000, Khasanov_HPR_2016, Khasanov_SST_2015} It also provides information on coexisting and competing phases in a material.\cite{Khasanov_SciRep_2015,  Khasanov_MnP_JPCM_2017, Khasanov_MnP_PRB_2016, Khasanov_Ni1144_PRB_2021, Khasanov_KCSRb122_PRB_2020, Gati_EuCdAs_PRB_2021, Gati_LaCrGe_PRB_2021} This is because muons stop uniformly throughout the sample, and the amplitude of the $\mu$SR signal arising from different regions of the sample are proportional to the volume occupied by each particular phase.

Transverse-field $\mu$SR is widely used to probe important length scales of
superconductors, namely the magnetic penetration depth $\lambda$ and the coherence length $\xi$.\cite{Smilga_book_1994, Lee_book_1999, Yaouanc_book_2011, Blundell_book_2022, Blundell_ContPhys_1999, Sonier_RMP_2000, Khasanov_HPR_2016, Khasanov_SST_2015} Experiments in the vortex state of a type II superconductor allow the determination of $\lambda$ in the bulk of the sample, in contrast to many techniques that probe $\lambda$ only near the surface. The penetration depth $\lambda$ is one of the fundamental parameters of a superconductor, since it is related to the density of superconducting carriers ($n_s$) and the supercarrier mass ($m^\ast$) by the London equation $1/\lambda^2 = \mu_0n_se^2/m^\ast$. The functional form of $n_s(T)$ depends on the gap symmetry and may bring direct information on the presence of multiple superconducting energy gaps.\cite{Khasanov_SST_2015, Khasanov_Bi-III_PRB_2018, Guguchia_npjQuantMat_2019, FonRohr_SciAdv_2019, Barbero_PRB_2018, Gupta_arxiv_2022}

The high sensitivity of $\mu$SR to internal fields makes it an extremely important technique for probing time reversal symmetry breaking (TRSB). Superconducting states with broken time reversal symmetry have Cooper pairs with non-zero magnetic moments.\cite{Luke_Nature_1998, Grinenko_SRO_NatCom_2021, Luke_PRL_1993, Maisuradze_PRB_2010, Shang_PRL_2018, Ghosh_JPCM_2021, Grinenko_NatPhys_2020, Grinenko_Ba122_NatPhys_2020} The TRSB charge density wave (CDW) order is characterized by carriers flowing in closed loops which also leads to the appearance of small moments. In both cases, the TRSB superconducting\cite{Luke_Nature_1998, Grinenko_SRO_NatCom_2021, Luke_PRL_1993, Maisuradze_PRB_2010, Shang_PRL_2018, Ghosh_JPCM_2021, Grinenko_NatPhys_2020, Grinenko_Ba122_NatPhys_2020} and TRSB CDW states,\cite{Mielke_Nature_2022, Guguchia_RbVSb_Arxiv_2022, Khasanov_CsVSb_PRR_2022, Yu_arxiv_2021, Smidman_Arxiv_2022} a local alignment of these moments results in extremely small internal magnetic fields which can be probed by the ZF-$\mu$SR technique.

In the following subsections, we present a few selected experimental results obtained by performing $\mu$SR experiments under pressure. For further details and examples, we refer to the original articles published during the last five years, Refs.~\onlinecite{Shermadini_HPR_2017, Naumov_PRA_2011, Khasanov_HPR_2022, Grinenko_SRO_NatCom_2021, Khasanov_Al_PRB_2021, Khasanov_Bi-II_2019, Khasanov_Bi-III_PRB_2018, Khasanov_Ga-II_PRB_2020, Khasanov_MnP_JPCM_2017, Khasanov_MnP_PRB_2016, Khasanov_SciRep_2015, Khasanov_KCSRb122_PRB_2020,  Das_PdBi2_PRL_2021, Gati_EuCdAs_PRB_2021, Gati_LaCrGe_PRB_2021, Guguchia_NatCom_2020, Guguchia_npjQuantMat_2019, Guguchia_MoTe2_JPSJ_2020, Guguchia_PRM_2019, Guguchia_LaBaCuO_PRB_2016, Hirashi_PRB_2020, Sugiyama_PRB_2020, Lamura_PRB_2020, Holenstein_PRB_2021, Holenstein_PRL_2019, FonRohr_SciAdv_2019, Zheng_PRB_2018, Simutis_PRB_2018, Khasanov_FeSe_PRB__2018, Khasanov_FeSe_PRB_2017, Barbero_PRB_2018, Majumder_PRB_2022, Majumder_PRL_2018, Forslund_SciRep_2019, Gupta_arxiv_2022, Guguchia_arxiv_2022, Forslund_arxiv_2022} and those mentioned in the previous review paper, Ref.~\onlinecite{Khasanov_HPR_2016}.

\subsection{The high-pressure magnetic state of MnP}

The pressure-temperature phase diagram of MnP was found to be rather complex. Starting from ambient pressure, MnP exhibits several magnetic transitions, with the last currently 'unknown` magnetic phase becoming a precursor to superconductivity.\cite{Cheng_PRL_2015} A superconducting dome with a maximum transition temperature of  $T_c\approx 1$~K emerges near a pressure of $p\simeq8$~GPa, where the long-range magnetic order vanishes.\cite{Cheng_PRL_2015}

\begin{figure}[htb]
\includegraphics[width=0.9\linewidth]{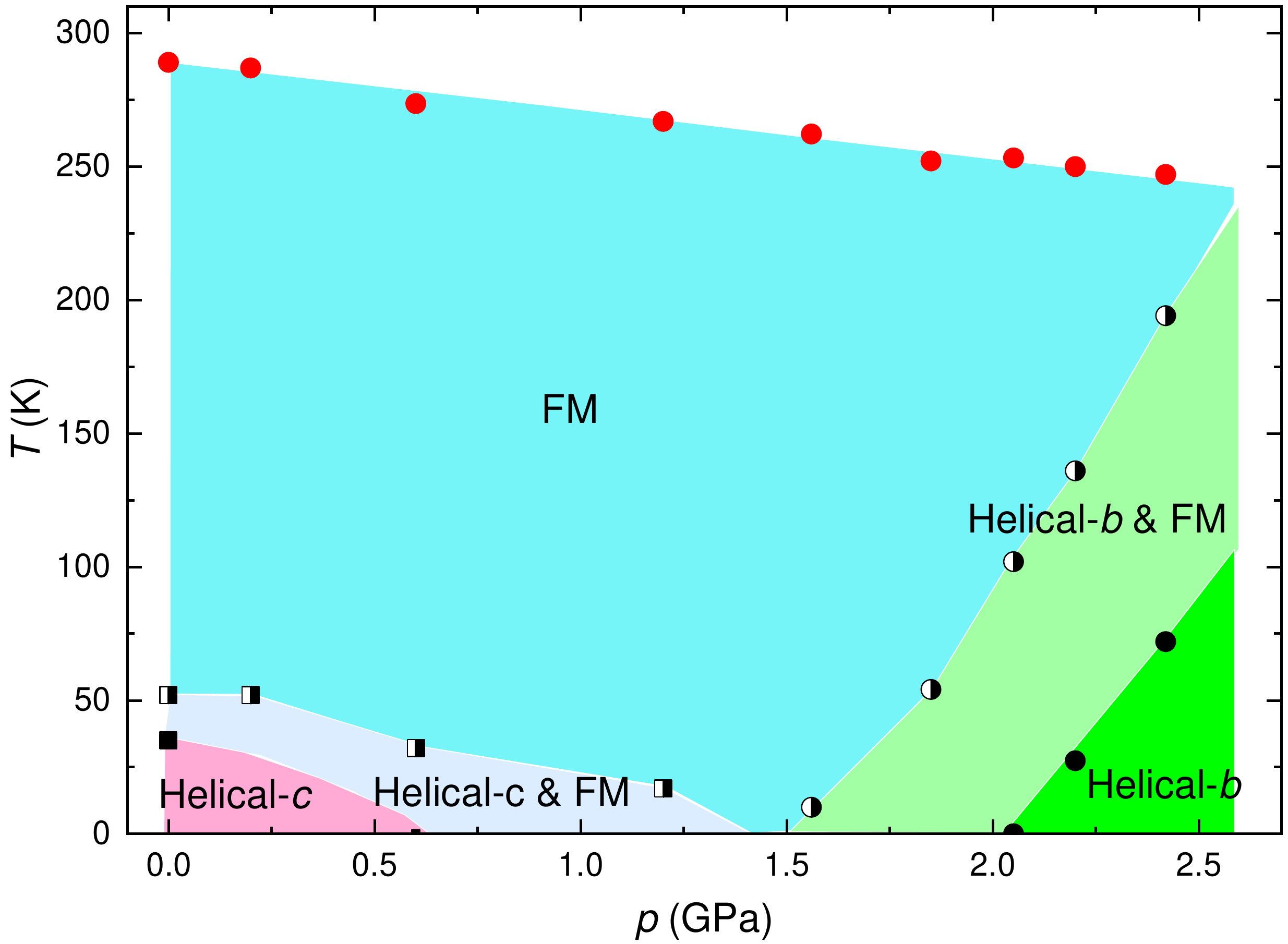}
%
\caption{(a) The $p-T$ phase diagram of MnP obtained from $\mu$SR experiments. FM, Helical$-c$, and Helical$-b$ denote different types of magnetic order (see the main text and Refs.~\onlinecite{Khasanov_MnP_PRB_2016, Khasanov_MnP_JPCM_2017}). The half filled and filled symbols correspond to the case when the helical$-c$ (pink squares) or the helical$-b$  (blue circles) phases occupy 10\% and 90\% of the sample volume, respectively. After Refs.~\onlinecite{Khasanov_MnP_PRB_2016, Khasanov_MnP_JPCM_2017}}
 \label{fig:Phase-Diagramm}
\end{figure}

In view of its ability to provide additional microscopic information on the magnetic and superconducting properties of MnP, ${\mu}$SR studies under applied hydrostatic pressure were conducted.\cite{Khasanov_MnP_PRB_2016, Khasanov_MnP_JPCM_2017} The resulting $p-T$ phase diagram  consists of three areas corresponding to the ferromagnetic (FM), helical$-c$  and helical$-b$  magnetic orders (see Fig.~\ref{fig:Phase-Diagramm}). The phases FM/Helical$-c$ as well as FM/Helical$-b$ are found to coexist within a broad range of pressures and temperatures. Transitions from the high-temperature FM to the low-temperature low-pressure helical$-c$, or the low-temperature high-pressure helical$-b$ phases are first-order like. These experiments appear to confirm that in MnP the high-pressure magnetic phase which is the precursor of the superconducting state is the incommensurate helical$-b$ type.

\subsection{Pressure Dependence of the Superfluid Density in the Nodeless Topological Superconductor $\alpha-$PdBi$_2$}

$\mu$SR experiments performed on topological superconductor candidate $\alpha-$PdBi$_2$ have suggested the absence of any nodes in the superconducting gap structure of this compound, and were found to be best described by a fully gapped $s-$wave model.\cite{Das_PdBi2_PRL_2021} Observation of $s-$wave superconductivity in this system along with its unique band structure and Fermi surface topology endorse this compound as a promising candidate in the search for Majorana zero modes. The superconducting gap value and the superfluid density were found to decrease with increasing pressure up to $p\simeq1.77$~GPa. The most intriguing observation of $\mu$SR studies under pressure is the linear dependence between $T_{\rm c}$ and the superfluid density manifested by the unconventional superconductivity of this compound.

\subsection{Unsplit superconducting and time-reversal symmetry breaking transitions in Sr$_2$RuO$_4$ under hydrostatic pressure and disorder.}

Sr$_2$RuO$_4$ was long considered a rare example of a $p_x \pm ip_y$ triplet chiral superconductor.\cite{Maeno_Nature_1994, Mackenzie_RMP_2003, Mackenzie_NPJ-Quantum_2017, Maeno_JPSJ_2012, Kallin_RepProgPhys_2012} Such a conclusion was based on the observation of a TRSB superconducting state in the ZF-$\mu$SR\cite{Luke_Nature_1998} and polar Kerr effect data,\cite{Xia_PRL_2006} as well as on the absence of suppression of the NMR Knight shift below the superconducting transition temperature.\cite{Ishida_Nature_1998} Chirality was also supported by the phenomenology of junctions between Sr$_2$RuO$_4$ and conventional superconductors providing evidence for the existence of domains within the superconducting state.\cite{Nakamura_JPSJ_2012, Anwar_SciRep_2013} However, triplet superconductivity conflicts with the strain dependence of the upper critical field,\cite{Steppke_Science_2017} and the recent revision of NMR data on unstrained Sr$_2$RuO$_4$.\cite{Pustogow_Nature_2019, Ishida_JPSJ_2019}

\begin{figure}[tbh]
\includegraphics[width=0.9\linewidth]{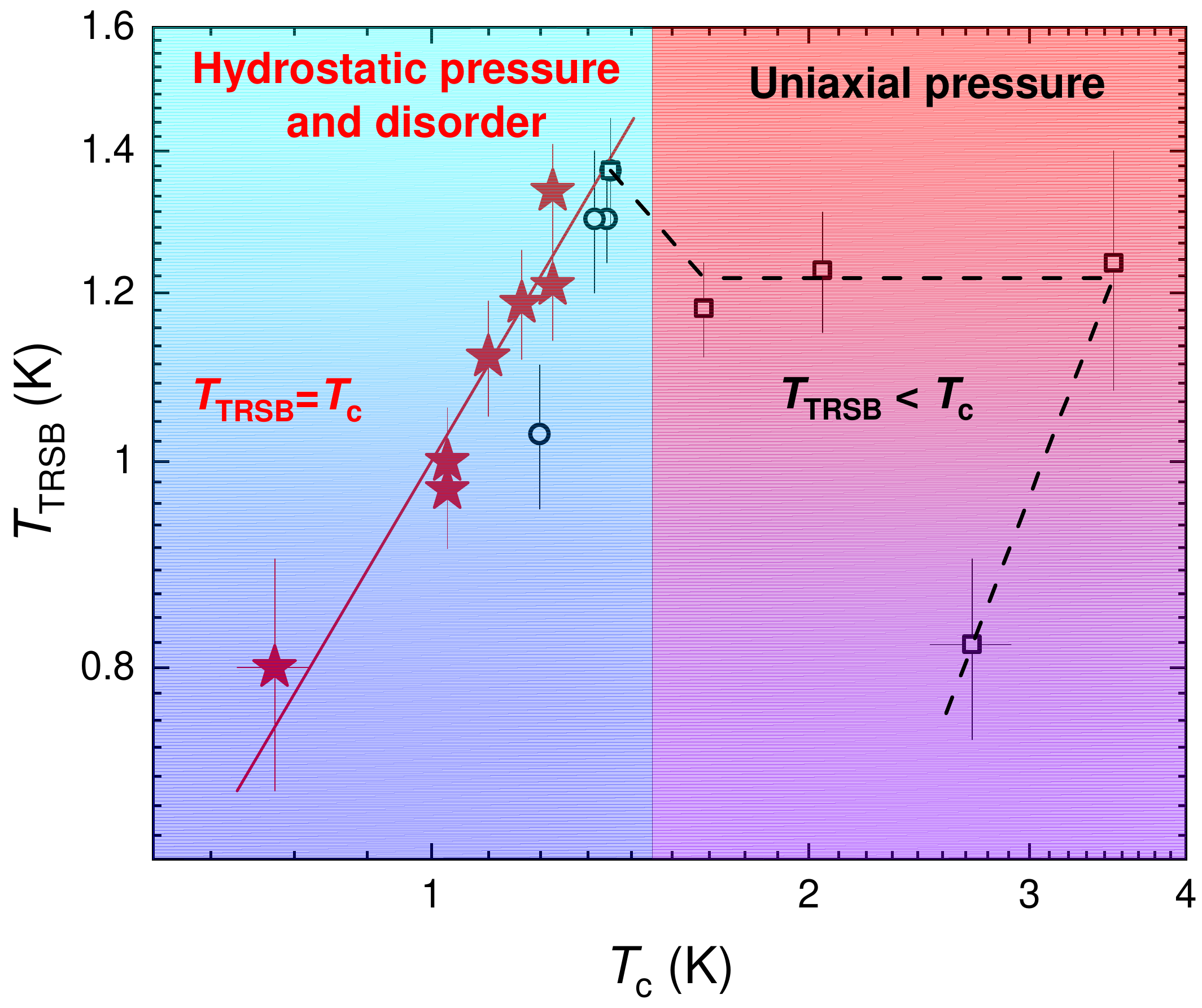}
\caption{Dependence of the time-reversal symmetry-breaking temperature $T_{\rm TRSB}$ on the superconducting transition temperature $T_{\rm c}$ for Sr$_2$RuO$_4$.\cite{Grinenko_SRO_NatCom_2021}.  The colored areas represents parts with preserved tetragonal lattice symmetry (hydrostatic pressure and disorder effects) and with orthorhombic distortions in the lattice (uniaxial pressure). The red solid line corresponds to $T_{\rm TRSB}=T_{\rm c}$. The dashed line is a guide to the eye.}
\label{fig:Tc-TTRSB_correlations}
\end{figure}

It was expected that hydrostatic pressure preserved the lattice-symmetry and disorder with an isotropic in-plane scattering potential that cannot split $T_{\rm c}$ and the TRSB transition temperature ($T_{\rm TRSB}$), while strong splitting between $T_{\rm c}$ and $T_{\rm TRSB}$ are expected for the perturbations breaking the crystallographic symmetry.
The recent $\mu$SR experiments reveal that the superconducting and TRSB transitions do not split under hydrostatic pressure up to 0.95~GPa in pure Sr$_2$RuO$_4$ and in the La doped Sr$_{2-x}$La$_x$RuO$_4$ sample with $T_{\rm c} \approx 0.7$K (see Fig.~\ref{fig:Tc-TTRSB_correlations}).\cite{Grinenko_NatPhys_2020, Grinenko_SRO_NatCom_2021} These results provide evidence in favor of chiral superconductivity in Sr$_2$RuO$_4$.

\subsection{Universal relations for type-I superconductors}

\begin{figure}[htb]
\includegraphics[width=0.85\linewidth]{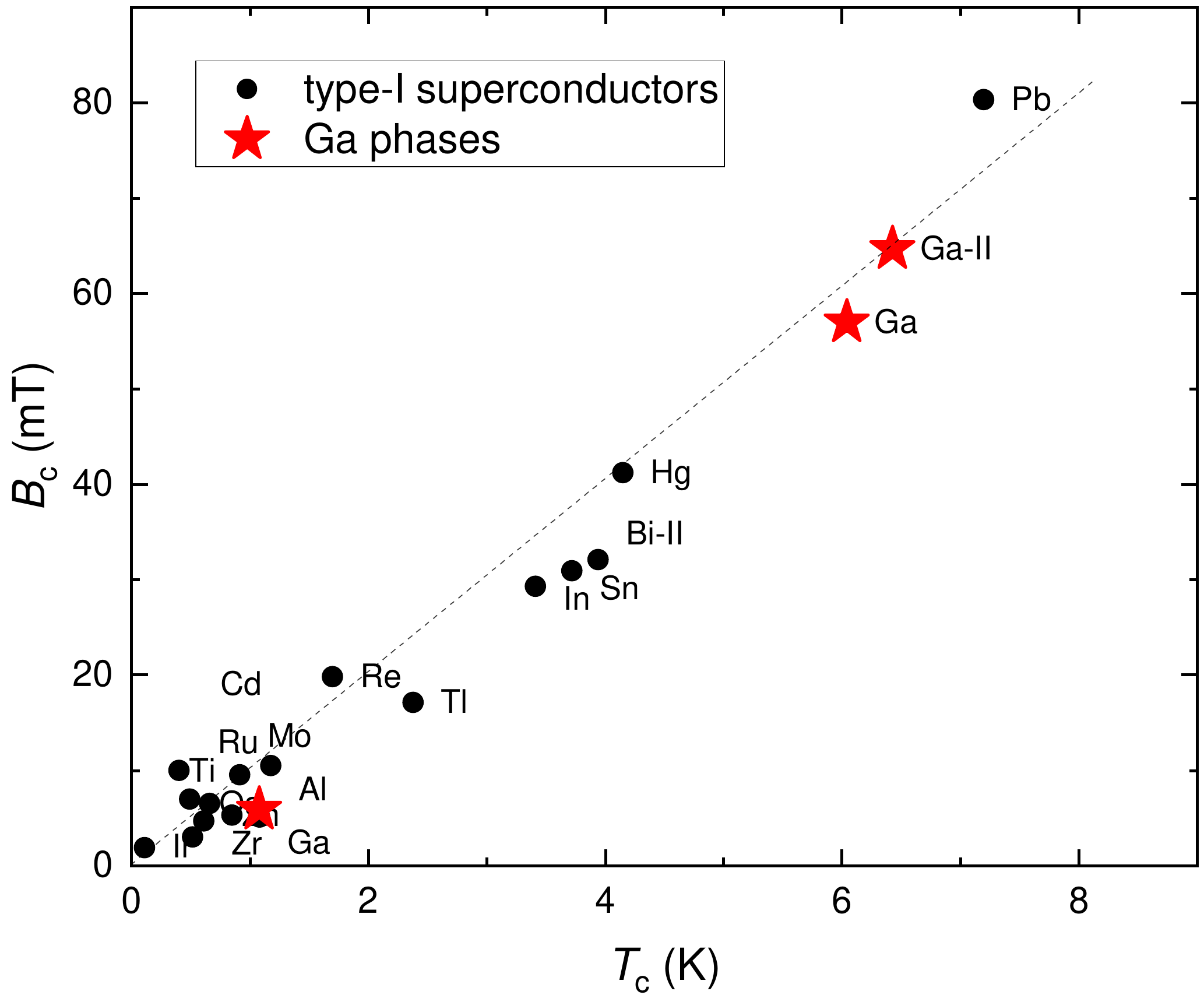}
\caption{The empirical relation between the zero-temperature value of the thermodynamic critical field $B_{\rm c}(0)$ and the transition temperature $T_{\rm c}$ for type-I superconductors, after Rohlf, Ref.~\onlinecite{Rohlf_Book_1994}. Black symbols are data points from Refs.~\onlinecite{Kittel_Book_1996, Poole_Book_2014, Leng_PRB_2017, Prakash_Science_2017, Campanini_PRB_2018, Beare_PRB_2019, Khasanov_Bi-II_2019, Karl_Sn_PRB_2019, Khasanov_AuBe_PRR_2020, Khasanov_AuBe_PRB_2020, Khasanov_Al_PRB_2021, Khasanov_Pb_PRB_2021}. Red stars correspond to different Gallium phases from Refs.~\onlinecite{Kittel_Book_1996,  Campanini_PRB_2018, Khasanov_Ga-II_PRB_2020}. After Ref.~\onlinecite{Khasanov_Ga-II_PRB_2020}.}
 \label{fig:Bc-vs-Tc}
\end{figure}

For type-I superconductors, the famous empirical trend reported by Rohlf in Ref.~\onlinecite{Rohlf_Book_1994} is the linear relation between the zero-temperature value of the thermodynamic critical field $B_{\rm c}(0)$ and the transition temperature $T_{\rm c}$, see Fig.~\ref{fig:Bc-vs-Tc}. The fact that the critical magnetic field required to destroy the superconducting state is strongly correlated with the critical temperature suggests that each of these parameters is representative of the energy supplied to the material that interferes with the superconducting mechanism. This is consistent with the idea that there is a bandgap between the superconducting and normal states. The proportionality between $T_{\rm c}$ and the superconducting gap is a well-accepted hypothesis, however it is not clear why $B_{\rm c}(0)$ should follow a similar trend.

With respect to the ability to detect superconducting responses, $\mu$SR studies of the pressure induced Gallium-II phase of the elemental gallium were conducted, Ref.~\onlinecite{Khasanov_Ga-II_PRB_2020}. The correlations between the coupling strength $2\Delta/k_{\rm B} T_{\rm c}$ ($\Delta$ is the superconducting gap), the ratio $B_{\rm c}(0)/T_{\rm c}\sqrt{\gamma_{\rm e}}$ ($\gamma_{\rm e}$ is the electronic specific heat), and the specific heat jump at $T_{\rm c}$, $\Delta C(T_{\rm c})/\gamma_{\rm e} T_{\rm c}$ has been shown to exist in phonon-mediated superconductors. Such correlations naturally explained the linear relation between $B_{\rm c}(0)$ and $T_{\rm c}$, and also provided an estimate of the $B_{\rm c}(0)/T_{\rm c}$ ratio.

\section{Outlook}

Muon-spin rotation/relaxation ($\mu$SR) measurements provide insight into how the muon interacts with its local environment. From this, unique information is obtained about the static and dynamic properties of the material of interest. This Perspective paper concentrates on the particular application of the $\mu$SR technique for conducting experiments under hydrostatic pressure conditions. The principal focus is on the description of the existing high-pressure muon facility at the Paul Scherrer Institute (PSI),  Switzerland.

Further possible development of PSI's $\mu$SR pressure facility has been foreseen in three important advances. First, a new fiber reading system is planned to be installed in several cryostats during the next few years. The use of such a system would allow one to monitor the pressure value within the full temperature range studied and, by doing so, allow a more accurate determination of the sample properties.  The second part relates to the improvement of the $\mu$SR pressure cells. This includes: (i) the search for new strong materials which might be suitable for building the $\mu$SR cells; (ii) the investigation of new pressure cell designs, such as, for example, McWhan-type cells, belt-type cells, three-wall cells, {\it etc.}; and (iii) the possible use of an anvil-type of design. The third part is intended for the development of an in-situ press system, which may allow one to change the pressure inside the cell directly at the muon instrument.

\begin{acknowledgments}
The author would like to thank Matthias Elender, Alex Amato, Elvezio Morenzoni, Hubetrus Luetkens, Zurab Guguchia, Charles Mielke III, Ritu Gupta, Debarchan Das, Kamil Sedlak, Gediminas Simutis, Hans-Ruedi Walter, Alexey Stoykov, Andrea Raselli, Kontstantin Kamenev, Stefan Klotz, Davide Reggiani, Thomas Rauber, Pavel Naumov, and Marek Bartkowiak  for their help and for making the $\mu$SR under pressure project to be running successfully.
\end{acknowledgments}

\end{document}